\newif\ifsubmode
\submodefalse

\newif\ifprintfig
\printfigtrue

\ifsubmode
  \documentstyle[12pt,aasms4,epsf]{article}
  \received{}
  \accepted{}
  \journalid{}{}
  \articleid{}{}
\else
  \documentstyle[11pt,aaspp4,epsf]{article}
  \slugcomment{{\it Astronomical Journal, December 2002, in press}}
\fi

\lefthead{van der Marel et al.}
\righthead{Central Kinematics of M15 --- I.~Observations}

\newcommand{\etal}{{et al.~}}

\newcommand{\lta}{\lesssim}

\newcommand{\kms}{\>{\rm km}\,{\rm s}^{-1}}

\newcommand{\RA}{{\rm RA}}
\newcommand{\DEC}{{\rm DEC}}

\newcommand{\hhh}{{\rm h}}
\newcommand{\mmm}{{\rm m}}
\newcommand{\sss}{{\rm s}}

\def\fr#1#2{{\textstyle {#1 \over #2}}}

\begin{document}

\title{Hubble Space Telescope Evidence for an Intermediate-Mass Black
Hole in the Globular Cluster M15---\\ 
I.~STIS Spectroscopy and WFPC2 Photometry\altaffilmark{1}}

\author{Roeland P.~van der Marel, Joris Gerssen}
\affil{Space Telescope Science Institute, 3700 San Martin Drive, 
       Baltimore, MD 21218}

\author{Puragra Guhathakurta, Ruth C.~Peterson\altaffilmark{2}}
\affil{UCO/Lick Observatory, Department of Astronomy and Astrophysics, 
       University of California at Santa Cruz, 1156 High Street,
       Santa Cruz, CA 95064}

\author{Karl Gebhardt}
\affil{Astronomy Department, University of Texas at Austin, Mail Code C1400,
       Austin, TX 78712}


\altaffiltext{1}{Based on observations made with the NASA/ESA Hubble Space 
Telescope, obtained at the Space Telescope Science Institute, which is
operated by the Association of Universities for Research in Astronomy,
Inc., under NASA contract NAS 5-26555. These observations are
associated with proposal \#8262.}

\altaffiltext{2}{Also at: Astrophysical Advances, Palo Alto, CA 94301.}


\ifsubmode\else
\clearpage\fi


\ifsubmode\else
\baselineskip=14pt
\fi


\begin{abstract}
In this series of two papers we describe a project with the Space
Telescope Imaging Spectrograph (STIS) on Hubble Space Telescope (HST)
to measure the line-of-sight velocities of stars in the central few
arcsec of the dense globular cluster M15. The main goal of this
project is to search for the possible presence of an intermediate mass
central black hole. This first paper focuses on the observations and
reduction of the data. We `scanned' the central region of M15
spectroscopically by consecutively placing the $0.1''$ HST/STIS slit
at 18 adjacent positions. The spectral pixel size exceeds the velocity
dispersion of M15. This puts the project at the limit of what is
feasible with STIS, and exceedingly careful and complicated data
reduction and analysis were required. We applied corrections for the
following effects: (a) drifts in the STIS wavelength scale during an
HST orbit; (b) the orbital velocity component of HST along the
line-of-sight to the cluster, and its variations during the HST orbit;
and (c) the apparent wavelength shift that is perceived for a star
that is not centered in the slit. The latter correction is
particularly complicated and requires many pieces of information: (1)
the positions and magnitudes of all the stars near the center of M15;
(2) the accurate positionings of the STIS slits during the
observations; (3) and the HST/STIS point-spread function (PSF) and
line-spread function (LSF). To address the first issue we created a
stellar catalog of M15 from the existing HST/WFPC2 data discussed
previously by Guhathakurta \etal (1996), but with an improved
astrometric and photometric calibration. The catalog is distributed
electronically as part of this paper. It contains 31,983 stars with
their positions and U, B and V magnitudes. To address the second issue
we model the observed intensity profiles along the STIS slits to
determine the slit positionings to $0.007''$ accuracy in each
coordinate. To address the third issue we obtained observations of a
bright field star to which we fitted multi-Gaussian PSF and LSF
models. Upon reduction of the M15 spectroscopy we ultimately obtain
19,200 one-dimensional STIS spectra, each for a different aperture
position in M15, with a velocity scale accurate to better than $2.5
\kms$. We develop an algorithm that co-adds the spectra for individual 
apertures and use it to extract spectra of individual stars with
minimum blending and maximum $S/N$. In Paper~II (Gerssen et al.) we
use these spectra to extract reliable line-of-sight velocities for 64
stars, half of which reside within $R = 2.4''$ from the cluster
center. These velocities constrain the central structure, dynamics and
mass distribution of the cluster.
\end{abstract}



\clearpage


\section{Introduction}
\label{s:intro}

This is the first in a series of two papers in which we present the
results of a study with the Hubble Space Telescope (HST) of the
line-of-sight velocities of stars in the central few arcsec of the
globular cluster M15 (NGC 7078). The present paper discusses the
spectroscopic observations with the HST Space Telescope Imaging
Spectrograph (STIS) and the extraction and calibration of the stellar
spectra. It also discusses the construction of a photometric catalog
from HST imaging with the Second Wide Field and Planetary Camera
(WFPC2), which proved essential in the reduction and interpretation of
the data. The second paper (Gerssen \etal 2002, hereafter Paper~II)
discusses the determination of line-of-sight velocities from the
spectra, and the implications for the dynamics and structure of
M15. The main motivation for our study is to better constrain the
possible presence of an intermediate-mass black hole in the center of
M15. Such a black hole was hinted at by previous work (e.g., Peterson,
Seitzer \& Cudworth 1989; Gebhardt \etal 2000a), but the limited
quality of the data precluded very strong conclusions (see the review
in van der Marel 2001).

The globular cluster M15 at a distance of 10 kpc has one of the
highest central densities of any globular cluster in our Galaxy. The
high density has likely affected the stellar population, as evidenced
by the presence of two bright X-ray sources (White \& Angelini 2001)
and several millisecond pulsars (Phinney 1993). A variety of
interesting dynamical and structural evolutionary phenomena have been
predicted to occur at high stellar densities, including mass
segregation and the possible formation of a central black hole (Hut
\etal 1992; Meylan \& Heggie 1997; see the additional discussion in
the introductory section of Paper II). As a result, M15 has been one
of the globular clusters for which the structure and dynamics have
been most intensively studied in the past decade.

Studies of the dynamics of M15 have focused primarily on the
determination of line-of-sight velocities of individual stars, using
either spectroscopy with single apertures, long-slits or fibers
(Peterson \etal 1989; Dubath \& Meylan 1994; Dull \etal 1997; Drukier
\etal 1998), or using imaging Fabry-Perot spectrophotometry (Gebhardt
\etal 1994, 1997, 2000a). Integrated light measurements using single
apertures have also been attempted (Peterson
\etal 1989; Dubath, Meylan \& Mayor 1994), but are only of limited
use; integrated light spectra are dominated by the light from only a
few bright giants, and as a result, inferred velocity dispersions are
dominated by shot noise (Zaggia \etal 1992, 1993; Dubath \etal 1994).

Line-of-sight velocities are now known for $\sim\! 1800$ M15 stars, as
compiled and analyzed by Gebhardt \etal (2000a). The projected
velocity dispersion profile increases inwards from $\sigma = 3 \pm 1
\kms$ at $R=7$ arcmin, to $\sigma = 11 \pm 1 \kms$ $R=24''$. The
velocity dispersion is approximately constant at smaller radii, and is
$\sigma = 11.7 \pm 2.8 \kms$ at the innermost available radius $R
\approx 1''$. The rotational properties of M15 are quite peculiar. The
position angle of the projected rotation axis in the central region is
$\sim\!  100^{\circ}$ different from that at larger radii, and the
rotation amplitude increases to $V_{\rm rot} = 10.4 \pm 2.7 \kms$ for
$R \lta 3.4''$, so that $V_{\rm rot} / \sigma \approx 1$ in this
region. A plausible explanation for these properties is still lacking.
 
To improve our understanding of the central mass distribution in M15,
it is of crucial importance to obtain more stellar velocities close to
the center. However, velocity determinations at $R \lta 2''$ are very
difficult due to severe crowding and the presence of a few bright
giants in the central arcsec. Gebhardt \etal (2000a) used an Imaging
Fabry-Perot spectrophotometer with adaptive optics on the CFHT, and
obtained FWHM values as small as $0.09''$. However, the Strehl ratio
was only $\lta 6$\%, so that even in these observations the light from
the fainter turnoff and main-sequence stars in the central arcsec was
overwhelmed by the PSF wings of the nearby giants. As a result, there
are only 5 stars with known velocities within $R \leq 1.3''$ from the
cluster center. To improve upon this situation we initiated the
project on which we report here, in which we used HST/STIS to map the
center of M15 spectroscopically at high spatial resolution. In the
future, additional kinematic data on M15 may come available from
stellar proper motion measurements with HST, using techniques such as
those described by Anderson \& King (2000).  However, at present no
such measurements are available.

This paper is organized as follows. Section~\ref{s:imaging} describes
the construction of a stellar catalog from WFPC2 data. This catalog is
essential for a proper interpretation of the M15 spectra. It is
distributed electronically as part of the present
paper. Section~\ref{s:calibspec} describes STIS spectra of a bright
field star which were obtained to calibrate the STIS point spread
function (PSF) and line-spread function (LSF). Section~\ref{s:specM15}
describes the M15 STIS spectroscopy, the observational setup and data
reduction, and an analysis of the achieved slit positioning
accuracies. The velocity calibration accuracy of the spectra is the
most crucial aspect of the observations, and this issue is discussed
in detail. Section~\ref{s:indstars} discusses the algorithm adopted
for the extraction of individual stellar spectra. These spectra form
the end-product of the present paper. They are used in Paper~II to
determine line-of-sight velocities and to study the structure and
dynamics of M15. Section~\ref{s:conc} presents concluding remarks.

\section{Construction of a Stellar Catalog}
\label{s:imaging}

For proper planning and interpretation of long-slit spectra of M15 it
is necessary to know the positions and magnitudes of the stars in M15.
Because M15 is dense, and its central regions are crowded, it is
important to determine these quantities from high spatial resolution
data. Several imaging studies of M15 were undertaken with the HST in
the years 1990-1994 (Lauer \etal 1991; Ferraro \& Paresce 1993;
Stetson 1994; de Marchi \& Paresce 1994; Yanny \etal 1994), but these
suffered from the aberrated HST PSF. Imaging studies with the Faint
Object Camera (FOC) on the refurbished HST (with COSTAR) were reported
by de Marchi \& Paresce (1996) and Sosin \& King (1997). However,
while FOC provided a sharp and well-sampled PSF, it had only a very
small field of view and its data were hard to calibrate
accurately. Data from WFPC2 are therefore preferable; see Biretta
\etal (2001) for a detailed description of this instrument.

\subsection{WFPC2 observations}
\label{ss:WFobs}

WFPC2 images of M15 were obtained on April 7, 1994 in the context of
HST program GO-5324 (PI: Yanny). Results from these observations were
reported in a study led by one of us (Guhathakurta \etal 1996,
hereafter G96). Data were obtained using the filters F336W, F439W and
F555W, respectively, which correspond roughly to the Johnson U, B and
V bands. The G96 paper provides a detailed discussion of the data
acquisition and reduction. Their figure~1 shows a large scale
black-and-white image of M15, while their figure~2 shows a
`true-color' image of the central $9'' \times 9''$. The photometric
analysis of G96 yielded a list of stars, and for each star its
position on one of the WFPC2 chips and the countrate in each of the
filters. This list is the starting point of the present analysis.
While G96 applied both an astrometric calibration (the transformation
from chip position to right ascension RA and declination DEC) and a
photometric calibration (the transformation from count-rate to Johnson
U, B and V magnitudes) to their star list, these were not particularly
accurate. The WFPC2 had only just been installed on HST when the G96
data were obtained, and there was little information available on the
appropriate astrometric and photometric transformation equations when
their paper was written. This was not a problem for the G96 study,
which focused on a determination of the stellar number density profile
of M15. However, for the present application both astrometric and
photometric accuracy are important, and we therefore decided to apply
an improved calibration to the G96 star list.

\subsection{Astrometric Calibration}
\label{ss:astrometric}

For the astrometric calibration we first ran the IRAF/STSDAS task {\tt
uchcoord}, as recommended by the HST Data Handbook (Voit \etal
1997). This properly sets the image header parameters for the WFPC2
plate scales and chip-to-chip rotations.  The task {\tt metric} was
then used to transform the pixel coordinates of each star to RA and
DEC in the Guide Star Coordinate System (GSCS). For the position of
the M15 cluster center we adopted the pixel position calculated by G96
(which has an estimated $1\sigma$ uncertainty of $0.2''$ in each
coordinate). For each star the relative position with respect to the
cluster center is given by $\Delta \RA = \RA - \RA_{\rm cen}$ and
$\Delta \DEC = \DEC - \DEC_{\rm cen}$. In the following it is more
convenient to work in a right-handed Cartesian coordinate system, so
we define $x = - \Delta \RA$ and $y = \Delta \DEC$. The 1$\sigma$
absolute accuracy of the GSCS is only $\sim 0.5''$. However, the
absolute astrometric accuracy can be improved using the star
AC211. This X-ray source has a radio counterpart for which the
position is determined to within $0.1''$ accuracy. Kulkarni \etal
(1990) find for AC211 that $\RA = 21\hhh \> 29\mmm \> 58.310\sss$ and
$\DEC = 12^{\circ} \> 10' \> 02.85''$. We find for AC211 that $x =
0.37''$ and $y = 1.96''$. This implies that the M15 cluster center is
at $\RA = 21\hhh \> 29\mmm \> 58.335\sss$ and $\DEC = 12^{\circ} \>
10' \> 00.89''$ in the radio coordinate system. This differs by
$0.55''$ from the position that was calculated in GSCS, consistent
with the absolute accuracy of the latter.

\subsection{Photometric Calibration}
\label{ss:photometric}

For the photometric calibration we started with the count rates (i.e.,
the integrated counts divided by the exposure time) determined by G96.
These had been corrected to an equivalent aperture with a radius of
$0.5''$, although the actual photometry was performed with a smaller
aperture.  These count rates were first corrected for geometric
distortion and then for the WFPC2 Charge Transfer Efficiency (CTE)
problem, as described in the HST Data Handbook (Voit \etal 1997). For
the CTE correction we used the formulae given by Whitmore \& Heyer
(1997). However, we increased their recommended corrections by a
factor 3, to take into account that the CTE was more of a problem
before April 24, 1994 (i.e., when the M15 data were taken). On this
date the WFPC2 temperature was lowered from $-76^{\circ}$C to
$-88^{\circ}$C. The factor of 3 was chosen on the basis of
quantitative comparisons of the CTE effect at the two different
temperatures (Holtzman \etal 1995a). We then transformed the count
rates to Johnson U, B and V magnitudes using the procedures described
by Holtzman \etal (1995b). Several iterations were performed to get
the color terms correct in the transformation equations. The filter-
and chip-dependent change in throughput that occurred on April 24,
1994 was taken into account, using the values given in the WFPC2 Data
Handbook (Voit \etal 1997).

The above steps provide an optimal relative photometric calibration
for the data. However, the absolute calibration remains uncertain due
to uncertainties in aperture corrections and contamination-related
throughput corrections. This is the case for WFPC2 data in general,
but is particularly important for data taken before April 24,
1994. WFPC2 was operated at $-76^{\circ}$C for only a few months, and
the instrument calibrations available for this period are considerably
less accurate than those available for subsequent data.  We therefore
applied a constant additive correction to all the stellar magnitudes
for a given filter+CCD combination. The corrections were chosen to
align the red giant branch (RGB) and horizontal branch (HB) features
in color-magnitude diagrams (CMDs) with the corresponding features in
well-calibrated ground-based data (Durrell \& Harris 1993; Stetson
1994). The corrections ranged from $0.1$--$0.3$ mag. With these
corrections we estimate the calibration of the final catalog to be
accurate to $\sim 0.03$~mag in $B$ and $V$ and to $\sim 0.05$~mag in
$U$. This is sufficient for the purposes of the present project.

\subsection{Properties of the Stellar Catalog}
\label{ss:catprops}

The final stellar catalog contains 31,983 stars with their positions
and U, B and V magnitudes. It is distributed electronically as part of
this paper. For reference, the first 25 entries are listed in
Table~\ref{t:catalog}. Each star has an ID number. These numbers were
already used for cross-identification purposes by Gebhardt \etal
(2000a) in their compilation of M15 stars with ground-based
line-of-sight velocity determinations. The spatial distribution of the
stars in our WFPC2 catalog is shown in Figure~\ref{f:catimage}. The
footprint of the WFPC2 field of view is clearly visible. Towards the
East the catalog extends as far as $2.0$ arcmin from the cluster
center. For comparison, the half-mass radius of M15 is $1.06$ arcmin
and the tidal radius is $22.1$ arcmin (Harris 1996), so the catalog
only covers the central part of M15. Note from Figure~\ref{f:catimage}
that stars are not detected along the CCD chip boundaries. Stars
within $14.9''$ from the cluster center are all detected on the PC
chip, and this inner region of M15 is therefore spatially complete in
the catalog.

Figure~\ref{f:CMD} shows a CMD of $V$ vs.~$B-V$ for the stars in the
catalog. The well-known stellar-evolutionary features of the main
sequence, subgiant branch, RGB, HB, and blue stragglers are all easily
discernible. The increased width of the HB between $0.2 \lta B-V \lta
0.5$ (the instability strip) is due to RR Lyrae variables. The
implications of the morphology and position of the observed CMD
features for the age, metallicity and stellar population of M15 have
all been addressed previously (e.g., Durrell \& Harris 1993; Ferraro
\& Paresce 1993; Stetson 1994; de Marchi \& Paresce 1994; Yanny \etal
1994; de Marchi \& Paresce 1996; Sosin \& King 1997) and will not be
discussed here. The large color spread of the main sequence at faint
magnitudes is due to photometric errors. The photometric errors and
completeness of the catalog, and their dependence on distance from the
cluster center, were discussed in detail by G96. The catalog is
essentially complete for magnitudes brighter than $V \approx 19$, and
the completeness drops steadily towards fainter magnitudes. No stars
are detected fainter than $V \approx 22.5$.

\section{STIS Calibration Spectroscopy of a Field Star}
\label{s:calibspec}

We were awarded 25 orbits in the context of HST program GO-8262 (PI:
van der Marel) for a spectroscopic study of the central region of M15.
All observations were taken with STIS (see Leitherer \etal 2001 for a
detailed description of this instrument). For all observations we used
the {\tt 52X0.1} slit, which has a calibrated width of $0.095''$
(Bohlin \& Hartig 1998). As dispersive element we used the G430M
grating centered at a wavelength of 5216{\AA}. This yields spectra
that cover the region from 5073---5359{\AA}. This wavelength range has
often been used for kinematical studies and it includes the Mg b
triplet region around 5175{\AA}. The pixel size in the spatial
direction is $0.05071''$ (Bowers \& Baum 1998). The pixel size in the
spectral direction is 0.276{\AA}. The latter corresponds to $15.86
\kms$ at the center of the wavelength range. For comparison, the
central velocity dispersion of M15 has previously been measured to be
$\sigma \approx 12 \kms$ (Gebhardt \etal 2000a), so for a proper
understanding of the M15 kinematics the Doppler shifts must be
measured to a fraction of a pixel accuracy. This requires accurate
calibration, and a detailed understanding of the STIS point spread
function (PSF) and line-spread function (LSF). For this reason we used
one orbit on June 14, 1999, for a set of calibration observations of a
bright field star.

\subsection{Observations}
\label{ss:calibobs}

For the calibration observations we chose to observe the star HD
122563, an extremely metal-poor field giant with magnitude $V=6.2$. At
the start of the orbit an {\tt ACQ} acquisition procedure was executed
to put the star near the center of the slit. This procedure has a
nominal accuracy between $0.01''$ (Leitherer \etal 2001) and $0.025''$
(Voit \etal 1997). It was not necessary to achieve better pointing
accuracy than this, so we did not perform an additional (more
accurate) {\tt ACQ/PEAK} procedure.  A series of small telescope slews
were then used to successively position the star at 14 different
positions relative to the slit, as shown schematically in
Figure~\ref{f:HDslit}. At each position two spectra were
obtained. These were dithered by an integer number of $\pm 2$ pixels
in the direction parallel to the slit, to allow removal of CCD defects
during data reduction. The exposure times ranged from 2 to 20 seconds
per exposure. The longer exposure times were used when the star was
offset perpendicularly from the center of the slit, so as to obtain an
approximately similar signal-to-noise ratio ($S/N$) for all
exposures. We used 2-pixel on-chip binning in the spatial direction to
obtain an effective pixel size of $0.1014''$.

The STIS Mode Selection Mechanism (which selects and positions the
grating) has a small mechanical non-repeatability. This causes an
uncertainty (amounting to a few pixels) in the relative position of
the 2D spectrum with respect to the CCD rows and columns. In addition,
slight drifts in this position (of the order of $0.1$ pixel) can occur
over the course of an orbit. For accurate calibration it is therefore
important that contemporaneous spectra of an arc lamp are
obtained. For the calibration star observations we obtained five
exposures of an arc lamp, spread equally in time between the beginning
and the end of the orbit.

\subsection{Data Reduction}
\label{ss:calibreduc}

Our data reduction procedures are based on the default STIS reduction
pipeline. However, we applied a number of modifications and additional
calibrations to obtain more accurate results. We started with the raw
data files, and used the IRAF/STSDAS task {\tt getref} to ensure that
the relevant header keywords point to the most appropriate calibration
files. Subsequently, the {\tt basic2d} task was run to perform the
tasks of data quality initialization, bias subtraction, dark current
subtraction, and flat-fielding. For the dark current subtraction we
used the `superdark' frames that are generated by STScI on a weekly
basis. We experimented with the use of daily darks, but found that
they did not improve the results due to their lower $S/N$. We fitted
and subtracted a spatially linear background from all frames to ensure
that source-free regions along the slit were consistent with zero. The
two exposures obtained at each stellar position were then aligned and
co-added. The latter was done with the task {\tt ocrrej}, which
rejects cosmic rays and CCD defects. Bad columns were corrected using
linear interpolation. After these steps a significant number of bad
pixels remain in the data. Positive outliers are generally pixels that
are `hot' in the data frame but normal in the (non-contemporaneous) dark
frame, while for negative outliers the opposite is generally the
case. All outliers were interpolated over using the task {\tt
cosmicrays} (to find the negative outliers the science frame was
multiplied by minus one).

The arc lamp spectra were reduced similarly as the object spectra.
The {\tt wavecal} task was then used to determine the shift of the
spectra with respect to a fiducial spectrum maintained in the HST Data
Archive. These Mode Selection Mechanism (MSM) shifts were interpolated
linearly in time over the course of the observations to determine the
MSM shift appropriate for each individual exposure of the target.

An additional complication in the velocity calibration arises from the
fact that HST moves around the Earth with an orbital velocity of $7.5
\kms$. This can cause velocity shifts between observations taken at 
different times of up to $15 \kms$. This exceeds the velocity
dispersion of M15, and it is therefore imperative to correct for
this. This is not normally done by the STIS/CCD pipeline. However, the
support files that are created and delivered with HST data contain
header keywords (VELOCSTX,VELOCSTY,VELOCSTZ) that contain the
three-dimensional velocity vector of HST (as measured halfway through
the exposure time). Phil Hodge (priv.~comm.) kindly provided a routine
to calculate the velocity component in the direction of the target,
given its RA and DEC. We used this routine to calculate this velocity
for each exposure.

For each exposure, the HST velocity shift was combined with the MSM
wavelength shift calculated previously (using the fact that 1 pixel is
$15.86 \kms$ at the center of the wavelength range). The shifts
calculated for different exposures at the same target position were
then averaged, weighted by the individual exposure times. These shifts
were then written to the header keywords in the co-added and reduced
target spectra. The task {\tt x2d} was then used to read these
keywords, and perform a two-dimensional rectification to a linear
heliocentric wavelength scale along rows, and a linear spatial scale
along columns. This also performs a flux calibration to units of ${\rm
erg} \> {\rm cm}^{-2} \> {\rm s}^{-1} \> {\rm \AA}^{-1}$.

\subsection{The Influence of Spatial Undersampling}
\label{ss:undersampling}

One problem in all STIS spectroscopy of point sources is the fact that
the spatial pixel size of $0.05071''$ undersamples the HST PSF. For
the calibration star observations we used 2-pixel on-chip binning in
the spatial direction to obtain an effective pixel size of $0.1014''$,
which further increases the undersampling. It is useful to have an
understanding of the influence of this on the inferred spectra. 

For any spectrograph, a spectrum that falls on the CCD is generally
slightly titled (by of order a few degrees) with respect to the
detector rows. For a point source that is well sampled, the 2D
spectrum will have the appearance of a smooth tilted line. However,
for a point source that is undersampled, the 2D spectrum will a have
the appearance of a stairwell. During data reduction, the spectrum is
interpolated and rotated to make it align with the rows of the frame.
However, this does not remove any undersampling. In an undersampled
situation, the rectified spectrum will on average be horizontal, but
it will have the zig-zag appearance of an un-tilted stairwell. If all
the rows in the rectified spectrum are summed, one recovers the
correct spectral shape. However, any horizontal cut along an
individual row will have strong low-frequency undulations multiplied
into the spectrum.

These effects are illustrated in Figure~\ref{f:undersamp}, which shows
the results for one of the two-dimensional spectra of the calibration
star. The top curve is the rectified spectrum summed along all rows,
which provides a fairly accurate representation of the actual spectrum
of the star. By contrast, the bottom curve shows the spectrum for the
brightest row in the rectified image. It has low-frequency undulations
multiplied into the spectrum. Individual rows in all of the rectified
spectra obtained for our program look similar to this. We obtained
most of the M15 data without spatial rebinning, but this does not
remove the undulations. It merely decreases their amplitude and
increases their frequency by a factor of 2.

At first glance, one might think that an undulating spectrum such as
in Figure~\ref{f:undersamp} is completely useless, unless one can
somehow correct for the undersampling. However, it turns out that for
the present purposes these undulations are not a problem at all. We
wish to measure radial velocities from the Doppler shifts in
absorption lines. The absorption lines are much narrower than the
undulations in the spectrum, and their positions are not in any way
affected. In cross-correlation techniques one always filters out low
frequency components, and we have found that this efficiently removes
any influence of the undulations on the inferred velocities.
Figure~\ref{f:crosscor} illustrates this by showing the
cross-correlation function of the two spectra shown in
Figure~\ref{f:undersamp}, both without and with low-frequency
filtering. The tests described below and in Paper~II have convinced us
that the undulations do not in any way compromise our ability to infer
radial velocities.

\subsection{Analysis}
\label{ss:calibanalysis}

After the reduction of the calibration star data, we have 14 different
two-dimensional spectra. In each spectrum there is signal in more than
one row, because the STIS PSF is not a delta function. All rows that
contain signal can be analyzed as separate spectra. For the analysis
it proved useful to also have a single grand-total one-dimensional
spectrum for each position of the calibration star. Such spectra were
obtained by co-adding all rows in a given two-dimensional frame.

We wish to use the spectra to calibrate the STIS PSF and LSF. For each
spectrum (either for a given row, or grand-total) we therefore
extracted two numbers: (a) the total count rate per second ${\cal C}$
summed over all wavelengths; and (b) the wavelength shift ${\cal S}$
of the spectrum with respect to some fixed standard spectrum. The
standard was chosen to be the grand-total spectrum obtained with the
star at the origin of the coordinate system shown in
Figure~\ref{f:HDslit}. The wavelength shifts were determined by
cross-correlation, using techniques that are described in detail in
Paper~II.

Figure~\ref{f:calibtot} shows ${\cal C}$ and ${\cal S}$ as function of
the offset $X$ of the star perpendicular to the slit (defined as in
Figure~\ref{f:HDslit}). The count-rate ${\cal C}$ was normalized
arbitrarily; its absolute value scales linearly with the brightness of
the star, and is of no interest in the present context. The count-rate
is large if the star is centered in the slit, and of course, it drops
if the star is moved further from the slit. However, the count-rate
remains positive even if the star is outside the slit, due to the
wings of the PSF. The peak of ${\cal C}$ identifies the center of the
slit, $X_{\rm slit}$. This peak does not occur at $X=0$, which implies
that the star was not perfectly centered in the slit after the target
acquisition procedure.

The shift ${\cal S}$ is measured in pixels (in the wavelength
direction). It is not a constant number, but shows a well-defined
behavior as function of $X$. The observed behavior is easily
understood qualitatively. At each wavelength, the grating makes an
image of the slit on the detector. If the light that enters through
the slit is not centered in the slit, then the observer perceives a
shift in wavelength. Let ${\cal S}_{\rm slit}$ be the shift (with
respect to the arbitrary standard) for a source placed in the center
of the slit ($X=X_{\rm slit}$). If the calibration star is positioned
at $X > X_{\rm slit}$, then the intensity-weighted average $X$
coordinate of the light that enters through the slit has $\langle X
\rangle > X_{\rm slit}$. This is perceived as a positive wavelength
shift, ${\cal S} - {\cal S}_{\rm slit} > 0$. Alternatively, if
$\langle X \rangle < X_{\rm slit}$, then ${\cal S} - {\cal S}_{\rm
slit} < 0$. If a star is moved out of the center of the slit, then
initially $|{\cal S} - {\cal S}_{\rm slit}|$ increases
monotonically. However, this behavior ultimately reverses. For large
values of $|X - X_{\rm slit}|$ only light in the wings of the PSF
enters through the slit. These wings have relatively small intensity
gradients, and the slit will be illuminated more or less
homogeneously. This causes $|{\cal S} - {\cal S}_{\rm slit}|$ to
revert to values near zero.

It is easy to see that $|\langle X \rangle - X_{\rm slit}|$ can never
be more than half the slit width, i.e., $0.0475''$ (this limit
corresponds to a delta function PSF). Taking into account the
anamorphic magnification of the grating, one detector pixel in the
wavelength direction corresponds to $0.05733''$ (Bowers
\& Baum 1998). Thus one always has $|{\cal S} - {\cal S}_{\rm slit}|
\leq 0.83$ pixels, or alternatively that the peak-to-peak variation in 
${\cal S}$ must be less than $1.66$ pixels. The observed peak-to-peak
variation in ${\cal S}$ is less, but still amounts to $0.84$ pixels
(cf.~Figure~\ref{f:calibtot}). This corresponds to $13.3 \kms$, which
is more than the M15 velocity dispersion. It is therefore important to
be able to model this effect, and to correct for it in the analysis of
the M15 data.

\subsection{PSF and LSF Model Fitting}
\label{ss:psflsf}

To model the calibration star data we consider the case of
observations of a point source through a rectangular aperture. The
aperture size is $A_x \times A_y$, where $A_x$ is the slit width and
$A_y$ the pixel size. We adopt a coordinate system with the origin at
the center of the aperture. We assume that the PSF can be modeled as a
circularly symmetric sum of Gaussians,
\begin{equation}
   {\rm PSF}(r) = \sum_{i=1}^N  \>
     {{\gamma_i}\over{2\pi \sigma_i^2}} \>
     \exp \left [- \fr{1}{2} ( \fr{r}{\sigma_i} )^2 \right ] .
\label{PSFGauss}
\end{equation}
The individual $\gamma_i$ may be either positive or negative, but must
satisfy $\sum_{i=1}^N {\gamma_i} = 1$. The associated encircled flux
curve $E(r)$ is given by
\begin{equation}
   E(r) \> \equiv \> \int_0^r {\rm PSF}(r') \> 2\pi r' \> dr' 
        \> = \> 1 - \sum_{i=1}^N {\gamma_i} \>
                \exp \left [- \fr{1}{2} ( \fr{r}{\sigma_i} )^2 \right ] .
\label{encircled}
\end{equation}
We assume that the aperture is perfectly rectangular, with no
diffraction off the aperture edges. Let the target be a point source
positioned at $(x_0,y_0)$. The fraction of the light from the point
source that is observed through the aperture is then
\begin{equation}
  f_{\rm tot} = 
     \sum_{i=1}^N \> {{\gamma_i}\over 4} \>
     \left \lbrace \mathop{\rm erf}\nolimits
            \left [ \fr{x_0+(A_x/2)}{\sqrt{2}\> \sigma_i} \right ]
           -\mathop{\rm erf}\nolimits
            \left [ \fr{x_0-(A_x/2)}{\sqrt{2}\> \sigma_i} \right ]
     \right \rbrace \>
     \left \lbrace \mathop{\rm erf}\nolimits
            \left [ \fr{y_0+(A_y/2)}{\sqrt{2}\> \sigma_i} \right ]
           -\mathop{\rm erf}\nolimits
            \left [ \fr{y_0-(A_y/2)}{\sqrt{2}\> \sigma_i} \right ]
     \right \rbrace ,
\label{PSFfrac}
\end{equation}
where ${\rm erf}(t)$ is the error function (van der Marel 1995). The
normalized intensity distribution in the $x$ direction, integrated
over the $y$-direction, of the light that enters through the aperture,
is
\begin{equation}
  f(x) = {1 \over {f_{\rm tot}}} \times \left\{ \begin{array}{ll}
     \sum_{i=1}^N \> 
       {{\gamma_i}\over{\sqrt{8\pi}\>\sigma_i}} \> 
       \exp[-{1\over 2}({{x_0-x}\over{\sigma_i}})^2] \>
       \left \lbrace \mathop{\rm erf}\nolimits
              \left [ {{y_0+(A_y/2)} \over {\sqrt{2}\> \sigma_i}} \right ]
             -\mathop{\rm erf}\nolimits
              \left [ {{y_0-(A_y/2)} \over {\sqrt{2}\> \sigma_i}} \right ]
       \right \rbrace ,
          & \quad \vert x \vert \leq {{A_x}\over2}        ; \\
     0 ,  & \quad \vert x \vert > {{A_x}\over2}           . 
                               \end{array} \right.
\label{fofx}
\end{equation}
For our observational setup, the coordinate $x$ is related to the
wavelength $\lambda$ through the equation ${\rm d}x / 0.05733'' = {\rm
d}\lambda / 0.276${\AA}. The LSF is the convolution of $f(x)$ with the
additional broadening functions due to the microroughness of the
grating, the natural diffraction properties of light, and the binning
into pixels (van der Marel, de Zeeuw \& Rix 1997). However, the latter
are all to lowest order symmetric functions of wavelength that do not
cause net wavelength shifts. The only net wavelength shift therefore
comes from the fact that $\langle x \rangle = \int x f(x) \> {\rm d}x$
(the intensity-weighted average $x$-coordinate of the light that
enters through the aperture) is generally different from zero (the
x-coordinate of the aperture center). Integration of
equation~(\ref{fofx}) yields
\begin{eqnarray}
  \langle x \rangle & = x_0 + 
       {1 \over {f_{\rm tot}}} \> \sum_{i=1}^N \> 
       {{\gamma_i}\over{\sqrt{8\pi}\>\sigma_i}} &
       \left \lbrace
           \exp \left  [ - \left ({{x_0+(A_x/2)} \over 
                                  {\sqrt{2}\> \sigma_i}} \right )^2
                \right ] -
           \exp \left  [ - \left ({{x_0-(A_x/2)} \over 
                                  {\sqrt{2}\> \sigma_i}} \right )^2
                \right ]
        \right \rbrace \nonumber \\ & & \times
       \left \lbrace \mathop{\rm erf}\nolimits
              \left [ {{y_0+(A_y/2)} \over {\sqrt{2}\> \sigma_i}} \right ]
             -\mathop{\rm erf}\nolimits
              \left [ {{y_0-(A_y/2)} \over {\sqrt{2}\> \sigma_i}} \right ]
       \right \rbrace . 
\label{xmean}
\end{eqnarray}
With the above equations we are now in a position to model the
observed quantities ${\cal C}$ and ${\cal S}$. The count-rate ${\cal
C} = K f_{\rm tot}$, where $K$ is a constant that is proportional to
the brightness of the target and the transmission of the
telescope+instrument system. The wavelength shift ${\cal S}$ is ${\cal
S} = 0.276{\rm \AA} \> (\langle x \rangle / 0.05733'')$, or alternatively
in terms of velocity, ${\cal S} = 15.86 \kms \> (\langle x \rangle /
0.05733'')$. The latter equation is formally valid only at the central
wavelength of the grating setting, but the dependence on wavelength
can be neglected in the present context.

One additional effect needs to be taken into account, namely the pixel
scattering function of the CCD. A photon that falls onto a given
detector pixel has a finite probability of being detected in an
adjacent pixel. This effect can be quantitatively described as
convolution with a kernel of size $3 \times 3$ pixels. The TinyTim
software package (Krist \& Hook 2001) provides an approximation for
this kernel at wavelengths of 4000{\AA} and 5500{\AA}. These can be
interpolated to the central wavelength of the our observations, which
yields: $(0.018, 0.052, 0.018 \> ; \> 0.052, 0.720, 0.052 \> ; \>
0.018, 0.052, 0.018)$. The convolution in the spectral direction of
the CCD corresponds to a symmetric line broadening, which does affect
our radial velocity measurements for M15. The only important effect in
the present context is the spatial broadening along the slit. This is
described by the 3 pixel convolution kernel $(0.088, 0.824, 0.088)$,
which is the spatial projection of the $3 \times 3$ kernel listed
above. To obtain predictions for the observed values ${\cal C}$ and
${\cal S}$ at a pixel $i$ along the slit, one must first calculate the
predictions at pixels $i-1$, $i$, and $i+1$, and then apply the
convolution kernel:
\begin{eqnarray}
  & {\cal C}_{i,{\rm obs}} & = 0.088 \> {\cal C}_{i-1} + 
                               0.824 \> {\cal C}_{i} +
                               0.088 \> {\cal C}_{i+1} ; \\
  & {\cal C}_{i,{\rm obs}}{\cal S}_{i,{\rm obs}} & = 
                               0.088 \> {\cal C}_{i-1} {\cal S}_{i-1} + 
                               0.824 \> {\cal C}_{i}   {\cal S}_{i} +
                               0.088 \> {\cal C}_{i+1} {\cal S}_{i+1}  .
\end{eqnarray}

To interpret the calibration star data we need to fit several
unknowns. The most important parameters to be determined are the
parameters $\gamma_i$ and $\sigma_i$ that describe the multi-Gaussian
PSF in equation~(\ref{PSFGauss}). In addition, there are: (a) the
constant $K$ defined through the equation ${\cal C} = K f_{\rm tot}$;
(b) the position $X_{\rm slit}$ of the slit center, which measures the
accuracy of the target acquisition procedure; (c) the shift ${\cal
S}_{\rm slit}$ (with respect to the arbitrary standard) for a source
placed in the center of the slit; and (d) the offset $\Delta Y$ of the
target with respect to the center of a pixel along the slit, at the
end of the target acquisition procedure. We assume that all telescope
slews between observations were accurate, as they are believed to be
(to the level of milli-arcsecs). The relative positions of the star
for the different observations are thus fixed by the values commanded
to the telescope (which are shown in Figure~\ref{f:HDslit}).

To constrain all the unknown parameters we used the measurements for
${\cal C}$ and ${\cal S}$, not only for the grand-total spectra as
shown in Figure~\ref{f:HDslit}, but also for each of the rows in the
individual two-dimensional spectra. While these constraints by
themselves were sufficient to obtain acceptable results, we 
found that the calibration star data do not very accurately constrain
the PSF at intermediate and large radii. To improve this situation we
used a model estimate for ${\rm PSF}(r)$ as an additional constraint
in the fit. For this, the TinyTim software was used to calculate the
STIS PSF at the central wavelength of our observations. While this PSF
is formally appropriate only for the STIS imaging mode, no better
approximation for spectroscopy is readily available. The TinyTim PSF
and its associated encircled flux curve are shown in
Figure~\ref{f:PSF}. Of course, addition of a model PSF as a constraint
in the fit can make the results worse rather than better, if the model
PSF is not in fact consistent with the calibration star data. We have
found no evidence for this, and in fact, from a variety of tests we
conclude that the two are fully consistent.

To determine the complete set of parameters in our multi-Gaussian PSF
model that best fits all the combined constraints, we used a `downhill
simplex' minimization routine (Press \etal 1992). We found that
adequate results were obtained with $N=5$ Gaussians in the
multi-Gaussian PSF. The thin solid curves in Figures~\ref{f:calibtot}
and~\ref{f:PSF} shows the predictions of the best fitting model with
this $N$. The parameters of the multi-Gaussian PSF are listed in
Table~\ref{t:psf}.

\subsection{Discussion}
\label{ss:calibdiscuss}

As has been discussed, there are many complications in the
determination of stellar velocities with HST/STIS at the few km/s
level: HST orbits the Earth as it exposes; the wavelength scale may
drift during an orbit; and a wavelength shift is perceived for a star
that is not perfectly centered in the slit. The most important result
of the calibration star observations is therefore the data-model
comparison shown in the bottom panel of Figure~\ref{f:calibtot}. The
model predictions fit the observed wavelength shifts ${\cal S}$ with
an RMS residual of $0.13$ pixels. This corresponds to $2.1 \kms$. So
if the position of a star with respect to the slit is known, then the
observed line of sight velocity can be calibrated to this accuracy.
Since the central velocity dispersion of M15 is $\sigma \approx 12
\kms$, this accuracy is sufficient for the purposes of the present
program. The excellent fit in the bottom panel of
Figure~\ref{f:calibtot} verifies not only our PSF and LSF modeling
procedures, but also provides a stringent test of our data acquisition
strategy and data reduction procedures.

Figure~\ref{f:PSF} shows that our multi-Gaussian PSF model is
essentially a smooth fit to the true PSF. Computational convenience is
gained, but some of the intricate detail of the PSF is lost. Most
noticeably, the multi-Gaussian PSF does not reproduce the first Airy
ring of the PSF (see in particular the middle panel of
Figure~\ref{f:PSF}). However, this is no reason for great concern. To
appreciate this, note that the PSF is itself a purely theoretical
construct that is generally unobservable. What is observed is the
convolution of the PSF with the response function of an individual
pixel, denoted the `effective PSF' (ePSF) by Anderson \& King
(2000). Due to the convolution with the pixel size, the Airy rings are
much less pronounced in the ePSF than in the PSF. As a result, the
ePSF of our model is much closer to the true ePSF, than is the PSF of
our model to true PSF. In fact, the observable quantities of interest
for this program are exactly those shown in Figure~\ref{f:calibtot},
and indeed, those are adequately reproduced by the model.

\section{STIS Spectroscopy of M15}
\label{s:specM15}

\subsection{Observations}
\label{ss:M15obs}

We obtained spectra of the central region of M15 for a period of 24
HST orbits, separated over 6 different visits. The first visit (5
orbits) was executed on October 21, 1999. Based on the analysis of the
data from the first visit some minor modifications were made to the
observing strategy. The remaining visits were then planned for the
third week of May 2001. Indeed, the data for visit 2 (4 orbits) were
obtained on May 15, 2001.  However, on May 16, 2001, the primary (Side
1) set of STIS electronics failed, which took the instrument out of
commission for two months. STIS was then revived using the redundant
(Side 2) set. This did not affect the instrument characteristics in a
major way, apart from a minor increase in the detector read-noise and
a somewhat different dark count rate. The remaining visits 3--5 (4
orbits each) and~6 (3 orbits) executed between October 22--26, 2001.

A complete log of the observations is presented in
Table~\ref{t:obslog}. All observations were done with the same slit
width and grating setting as described in Section~\ref{s:calibspec}.
Exposures were taken at 18 different positions covering the central
region of the cluster. Figure~\ref{f:center} shows the slit positions
as they were commanded to the telescope. The accuracy with which the
intended positioning was achieved is discussed in
Section~\ref{ss:slitpos} below.  For all observations the slit was
aligned along position angle ${\rm PA} = 26.65^{\circ}$ (measured from
North over East). This PA was chosen to allow the most flexible
scheduling of the observations, and is otherwise fairly arbitrary. At
each slit position we obtained spectra for a total exposure time that
ranged from 46 to 60 minutes. Three or four exposures were obtained
per position. Individual exposures were dithered by 4 pixels along the
slit, to allow removal of CCD defects during data reduction. In
visit~1 we placed the central region of M15 near the center of the
CCD. In subsequent visits we placed the central region near row 900
(out of 1024), to mitigate the loss of signal due to deteriorations in
the CTE of the STIS/CCD. In no case did we apply on-chip rebinning in
the spectral direction, given that the $15.86 \kms$ size of a single
pixel is already large compared to the M15 velocity
dispersion. However, in the first visit we applied 2-pixel on-chip
rebinning in the spatial direction (as for the calibration star
observations discussed in Section~\ref{s:calibspec}) to minimize the
read-noise component in the spectra. In subsequent visits we did not
apply such on-chip rebinning, to minimize the effects of PSF
undersampling in the spatial direction. The large majority of the
orbits consisted of the same simple observational sequence: an arc
lamp exposure, followed by three target exposures and then another arc
lamp exposure. In the end, the spectra from all visits had sufficient
quality to achieve the science goals of our program.

Blind pointings with HST are accurate only at the level of $\sim
0.5''$, due to the accuracy of the GSCS. To position the STIS slit on
M15 with the required accuracy it is therefore necessary to perform a
target acquisition. STIS target acquisitions work by centering on the
brightest object in a $5'' \times 5''$ field. To ensure that the
target acquisition would not acquire the wrong target, we searched for
a bright and relatively isolated star in M15. The cross in
Figure~\ref{f:catimage} marks the spatial position of the star that
was chosen (\#27604 in the WFPC2 catalog). Its position in the $(x,y)$
coordinate system defined in Section~\ref{ss:astrometric} is
$(-3.286'',-39.344'')$. Its position in the M15 CMD shown in
Figure~\ref{f:CMD} is also indicated by a cross. Each visit started
with an acquisition of this star, consisting of an {\tt ACQ}
acquisition procedure. In most visits we then slewed the telescope to
the intended position near the center of M15 to take the
spectra. However, in visits~1 and~6 we performed an additional {\tt
ACQ/PEAK} acquisition procedure (Leitherer \etal 2001) and obtained
spectra of the acquisition star, before slewing to the center of M15.
In visit~1 we also repeated this procedure at the end of the visit, to
allow an independent check of the repeatability of the velocity
measurements for this particular star. The positional accuracy of the
target acquisitions and the slews is discussed in
Section~\ref{ss:slitpos} below.  We generally performed a target
acquisition only at the beginning of the first orbit of each visit,
and not in any of the subsequent orbits.

\subsection{Data Reduction}
\label{ss:M15reduc}

The data reduction of the M15 spectra was similar to the reduction of
the calibration star spectra discussed in Section~\ref{ss:calibreduc},
including the wavelength calibration using arc lamp spectra obtained
in the same orbit, and the correction for the HST velocity during the
observations. This yields for each slit position a single
two-dimensional frame with a linear heliocentric wavelength scale
along rows, and a linear spatial scale along columns.

One difference between the observations of M15 and those of the
calibration star is that we obtained only two arc-lamp spectra per
orbit for M15, as compared to five for the calibration star. We did
some quantitative analysis to address to what extent this may affect
the analysis. The arc lamp spectra show that there are drifts in the
wavelength scale over the course of a visit. These drifts are
generally smoothly varying, and are fairly well modeled as a linear
function of time. In the worst case (visit~2) the drift was $1.3$
pixels over the 4 orbits of the visit. Because we do in fact take two
arc-lamp spectra per orbit, we can correct for most of this. In the
reduction of the M15 spectra we assume that the wavelength scale for a
given exposure can be obtained by linear interpolation in time between
the results for the nearest arc lamp spectra. To test the accuracy of
this, we applied this same procedure to each of the arc-lamp
spectra. This yields the difference between the actual wavelength
scale and the one that is obtained by linear interpolation between the
previous and the next arc-lamp exposure in the visit. The RMS
difference for all arc lamp spectra was found to be $0.08$ pixels,
which corresponds to only $1.3 \kms$. This accuracy is sufficient for
the purposes of the project, especially since other effects already
set a lower limit to the velocity accuracy of $2.1 \kms$
(cf.~Section~\ref{ss:calibdiscuss}).

\subsection{Slit Positioning Accuracy}
\label{ss:slitpos}

For accurate velocity determination of an M15 star it is essential to
know its position with respect to the slit. From the WFPC2 imaging we
know where all the stars are on the sky
(see~Section~\ref{s:imaging}). So the next step is to determine the
exact positioning of the slits for all observations. For this we
analyzed and modeled the observed intensity profiles along the
slit. These essentially provide one-dimensional (linear) images of
M15, which uniquely constrain the telescope pointing.

To model an observed intensity profile along the slit we proceed as
follows. For each star in the WFPC2 catalog we use the observed
Johnson B and V magnitudes to estimate the flux $F_{\nu}$ at 5216{\AA}
(the center of the wavelength range for the spectra). We choose a
trial position of the slit on the sky. For each pixel along the slit
we calculate the flux contributed by each star, using the formulae
derived in Section~\ref{ss:psflsf}. Adding the contributions of all
stars in the catalog yields the predicted intensity profile along the
slit for the given assumed slit position. This profile can be compared
to the observed profile in a $\chi^2$ sense (using an arbitrary
scaling that corresponds to the efficiency of the telescope+instrument
system). Looping over a fine grid of possible slit positions yields
the position with minimum $\chi^2$, which corresponds to the estimated
position of the slit on the sky. To obtain adequate fits we found that
we had to add a smooth, low-intensity model for the light contribution
of faint stars that are missing from the WFPC2 catalog due to
incompleteness. As an example of the procedure, Figure~\ref{f:intprof}
shows the final data-model comparison for one of the slit positions.

The accuracy with which HST can achieve a requested roll angle is
$\sim 0.003^{\circ}$. The accuracy with which the slit PA on the sky
is known is dominated by the calibration uncertainty in the relative
angle of the slit within the instrument, which is $\sim
0.05^{\circ}$. This is small enough that it can be neglected in the
present context. We therefore did not include the slit PA as a free
parameter in the fits to the intensity profiles, but kept it fixed at
the value commanded to the telescope. To confirm the validity of this
approach we did a number of test runs in which we fitted observed M15
intensity profiles with the slit PA treated as a free parameter. These
tests indeed implied PA values consistent with the value commanded to
the telescope, to within the quoted uncertainty.

The intensity profile modeling yields for each observation the actual
pointing position of the telescope during the exposures. The intended
pointings are known as well, so it is possible to determine for each
observation the offset $(\Delta X,\Delta Y)$ of the actual pointing
from the intended pointing. The results are shown in
Figure~\ref{f:pointings}. Observations from the same visit are shown
with the same plot symbol. The offsets satisfy $|\Delta X| < 0.1''$
and $|\Delta Y| < 0.1''$ for all observations. These absolute pointing
accuracies are not so important in the present context; the aim of the
observational strategy was to scan the central region of M15, and not
to point at any object in particular. However, it is important to know
the accuracy of our determinations of $(\Delta X,\Delta Y)$. To this
end we need to understand the sources of the inferred pointing
offsets.

The first source of pointing error is the limited astrometric accuracy
of the WFPC2 catalog. Figure~\ref{f:catimage} shows that the star that
was used for target acquisitions was on a different CCD chip in the
WFPC2 observations than the center of M15. The astrometric calibration
of WFPC2 is considerably less accurate across CCD chips than within a
given CCD chip ($0.1''$ as compared to $0.005''$, according to the HST
Data Handbook; Voit \etal 1997). Since the slew that we applied from
the acquisition star to the M15 center was based on the astrometry of
our catalog, an error in this astrometry would yield a fixed non-zero
component to the offsets $(\Delta X,\Delta Y)$ for all observations.

The second source of pointing error is the limited accuracy with which
the positions of the apertures are known within the STIS instrument.
An {\tt ACQ} acquisition centers the target in the {\tt 0.2X0.2}
aperture. On-board software then applies a telescope slew to place the
target behind the science aperture (the {\tt 52X0.1} in the present
case). If there is an error in the expected position of the science
aperture, this yields a fixed non-zero pointing error relative to the
slit. We obtained the data for visit~2 in a configuration that is
rotated by $180^{\circ}$ with respect to that for the other visits. So
any component in the offsets $(\Delta X,\Delta Y)$ due to this effect
would have the opposite sign for visit~2 as compared to the other
visits. Figure~\ref{f:catimage} shows that the average offset for
visit~2 is $(-0.059'',0.004'')$, whereas the average for all the other
visits is $(0.066'',0.024'')$. So there is indeed evidence for
pointing errors of this nature.

The third source of pointing error is the limited accuracy of the
target acquisitions that start each visit. For most visits we only
used an {\tt ACQ} acquisition procedure, and not the more accurate
{\tt ACQ/PEAK} procedure. The former has a nominal accuracy between
$0.01''$ (Leitherer \etal 2001) and $0.025''$ (Voit \etal 1997).

The fourth source of pointing error is the limited accuracy of the
$39.5''$ slews from the acquisition star to the M15 center. The
nominal accuracy of such slews is $0.02''$ (Downes
\& Rose 2001). 

The first four sources of error can explain both the size and the
scatter in the pointing offsets for different visits. However, these
error sources predict identical offsets for all observations in the
same visit. In reality, there is some scatter among the pointing
offsets for observations in the same visit, but this scatter is very
small: $0.005''$ in the $X$ direction, and $0.007''$ in the $Y$
direction. We attribute this to the fifth source of pointing error:
small pointing drifts during a visit as a result of thermal effects on
the telescope (Downes \& Rose 2001).

The small scatter in the offsets for observations in the same visit
implies that our models for determining the slit positions from the
data must be very accurate. The numerical accuracy can be no worse
than $0.007''$ in each coordinate. An error of this size in the STIS
$X$ coordinate corresponds to a wavelength shift $\Delta {\cal S}$ of
at most $0.08$ pixels, cf.~the bottom panel of
Figure~\ref{f:calibtot}. This corresponds to $1.2
\kms$. This accuracy is sufficient for the purposes of the project, 
especially since other effects already set a lower limit to the
velocity accuracy of $2.1 \kms$ (cf.~Section~\ref{ss:calibdiscuss}).

\section{Extraction of Spectra for Individual Stars}
\label{s:indstars}

Each row in each of the 18 calibrated and rectified two-dimensional
spectra contains the spectrum for a small aperture on the sky. From
the analysis in Section~\ref{ss:slitpos}, the positions of these
apertures are known with a (two-dimensional) accuracy better than
$0.01''$. Figure~\ref{f:apertures} shows the inferred aperture
positionings for the central region of M15. In total there are 19,200
spectra in the dataset. Most spectra contain very little signal from
any star, and are therefore of little use in the analysis. However,
some spectra do contain considerable signal from one or more M15
stars. Many stars contribute significantly to more than one
aperture. For proper analysis we therefore developed an algorithm to
extract the spectra of individual stars.

The algorithm starts with the same modeling procedure used in
Sections~\ref{ss:psflsf} and~\ref{ss:slitpos}. From the known
positions of all apertures $i$ and all stars $j$ we calculate
predictions for three sets of quantities: (i) $F_i$, the total flux
observed in an aperture $i$; (ii) $F_{ij}$, the total flux contributed
by star $j$ to aperture $i$; and (iii) ${\cal S}_{ij}$, the velocity
offset with which star $j$ contributes flux to aperture $i$. Let $F_j$
be the total flux of star $j$. We define $g_{ij} = F_{ij} / F_j$ to be
the fraction of the flux of star $j$ that falls in aperture $i$. We
also define $f_{ij} \equiv F_{ij} / F_i$ to be the fraction of the
total flux in aperture $i$ that is contributed by star $j$. For each
aperture $i$ there is one star $j$ in the catalog for which $f_{ij}$
has the largest value. We denote the index number of this star as
$J(i)$. We denote an aperture as `usable' if the following three
statements are true: (i) $f_{iJ(i)} \geq 0.5$ (i.e., there is a single
star that contributes at least half of all the flux in the aperture);
(ii) $F_{iJ(i)} \geq 6.3 \times 10^{-18} \> {\rm erg} \> {\rm cm}^{-2}
\> {\rm s}^{-1} \> {\rm \AA}^{-1}$ (i.e., there is a non-negligible
amount of flux from this star in the aperture; this limit corresponds
to a $V$-band magnitude of $\sim 22$); and (iii) $g_{iJ(i)} \geq
0.005$ (the aperture contains a non-negligible fraction of the total
flux of this star). Apertures that are not usable are ignored in all
of the subsequent analysis. For each star $j$ we then make a list of
all the apertures $i$ for which $J(i) = j$. Let $N(j)$ be the number
of apertures for which this is the case. We denote a star to be
`significant' for our study if $N(j) > 0$. We call the spectra for the
usable apertures that make up $N(j)$ the `usable' spectra for star
$j$.

In the full dataset there are 1741 usable spectra for 636 significant
stars. The goal is to combine the $N(j)$ usable spectra for each
significant star $j$ into a final spectrum for that star. This can be
done by taking the exposure-time weighted average of the usable
spectra. However, there is the possibility to include or exclude each
usable spectrum in or from the average, so there are $2^{N(j)}$
possible combinations to do the averaging. Specifically, one does not
want to include spectra whose main effect is to increase the amount of
noise or blending in the average. We developed an algorithm to
optimize the quality of the final spectrum. The STIS data reduction
pipeline creates an error frame for each spectrum. We use the data
frames and the error frames to calculate for each usable spectrum the
median flux $Z_{ij}$ from the star (i.e., the median observed flux
multiplied by $f_{ij}$) and the median noise. From these quantities we
calculate the signal-to-noise ratio $S/N$ for all of the $2^{N(j)}$
possible combinations of spectra. We also calculate for all
combinations the fraction $h$ of the total flux in the averaged
spectrum that is due to the significant star. We rejected all
combinations in which $h$ is smaller than some value $h_{\rm min}$.
Of the remaining combinations we adopt the one that produces the
highest $S/N$. The instrumental velocity shift ${\cal S}_{j}$ of the
final average spectrum of star $j$ is calculated as the weighted
average of the ${\cal S}_{ij}$ for the apertures $i$ that are used in
the combination; the weighting is done using the product of exposure
time and the flux $Z_{ij}$.

We adopted $h_{\rm min} = 0.75$ in the extraction, to ensure that the
fraction of the flux in the final spectrum that is blended in from
other stars is less than 25\%. For the spectra to be useful for
kinematical analysis they must have a sufficient $S/N$. As
demonstrated in Paper~II, an average $S/N \geq 5.5$ per pixel is a
minimum requirement (although not a sufficient requirement) to obtain
meaningful line-of-sight velocity measurements. All extracted spectra
that did not satisfy this requirement were discarded. This ultimately
yielded `acceptable' spectra for 131 stars, out of which 15 stars are
within $1''$ from the cluster center, 39 stars are within $2''$, and
87 stars are within $5''$.

Figure~\ref{f:spectra} shows the spectra for four stars in the final
sample. All four spectra correspond to stars on the giant branch. From
top to bottom, the $V$ magnitudes range from $V=13.74$ to $V=18.10$
while the average $S/N$ per pixel ranges from $126.1$ to
$7.3$. Although the bottom spectrum is quite noisy, the Mg b triplet
lines at $\sim 5170${\AA} are still clearly visible. The faintest star
for which we were able to extract a spectrum with $S/N > 5.5$ from the
STIS data has magnitude $V = 19.15$, which is just above the main
sequence turnoff.

The spectra in Figure~\ref{f:spectra} are all reasonably `flat', and
do not show the undulations seen in Figure~\ref{f:undersamp}. This is
because the spectra are sums of individual apertures for several slit
positions.  Each of the individual spectra has undulations, but these
tend to cancel in the final co-addition. Approximately 10 per cent of
the stars in our final list of 131 are based on only a single
aperture. The final spectra for these stars show undulations similar
to those in Figure~\ref{f:undersamp}. As explained in
Section~\ref{ss:undersampling}, this does not compromise our ability
to infer radial velocities for these stars.

The 131 individual stellar spectra that we extracted from the STIS
data form the end product of the present paper. It is shown in Paper
II that accurate velocities can be determined for 64 of these
stars. The spatial and color-magnitude distributions of these stars
are discussed in Paper II. That paper also discussed the implications
of the inferred velocities for our understanding of the structure and
kinematics of M15.

\section{Conclusions}
\label{s:conc}

The globular cluster M15 is one of the densest globular clusters in
our Galaxy. A variety of interesting dynamical and structural
evolutionary phenomena have been predicted to occur at high stellar
densities, including mass segregation and the possible formation of a
central black hole. Stellar kinematical data are required required to
address these issues observationally. While M15 has been one of the
globular clusters that has been most intensively studied in the past
decades, ground-based data have only been able to provide limited
information in the central few arcsec. We therefore executed a project
to map the center of M15 spectroscopically at high spatial resolution
with HST/STIS.

Our observational project is at the limit of what is feasible with
HST/STIS. The pixel size of our spectra corresponds to $15.86 \kms$,
which exceeds the central velocity dispersion of M15 ($\sim 12 \kms$;
Gebhardt \etal 2000a). Extreme care therefore had to be taken in the
reduction and velocity calibration of the data. The main result of the
present paper is that we have in fact succeeded in extracting
accurately calibrated spectra from our data. The implications of this
for our understanding of M15 are discussed in Paper II.

To be able to properly interpret the long-slit spectra, we started by
creating a stellar catalog for M15. We used the HST/WFPC2 data
discussed previously by G96, but we performed an improved astrometric
and photometric calibration. The final stellar catalog contains 31,983
stars with their positions and U, B and V magnitudes. It is spatially
complete within $14.9''$ from the cluster center, and extends as far
as $2.0$ arcmin from the cluster center in some directions. The
catalog contains stars down to magnitude $V \approx 22.5$, and it is
photometrically complete for stars brighter than $V \approx 19$.

For proper calibration of the M15 spectra we performed a set of
observations of a bright field star, with the star placed at various
positions with respect to the slit. After basic data reduction and
correction for the velocity of HST around the Earth we used these
spectra for modeling the PSF and LSF of STIS. This is particularly
important for understanding the apparent velocity shift for a star
that is offset from the center of the slit. We derive a simple
multi-Gaussian model for the PSF that reproduces the observed shifts
with an RMS residual of only $2.1 \kms$ (for a known stellar position
with respect to the slit). Spatial undersampling causes artificial
undulations in the spectra, but this only affects low frequencies and
does not affect the determination of radial velocities through
cross-correlation.

The M15 observations consisted of 6 HST visits spread over 2 years.
The central region of the cluster was `scanned' by consecutively
placing the $0.1''$ slit at 18 adjacent positions. The intensity
profiles along the slits were used to accurately determine the actual
slit positions on the sky. With use of our WFPC2 catalog and PSF model
we find that these positions can be determined to an accuracy of
$0.007''$ in each coordinate. Positional errors in our knowledge of
the slit positions translate directly into velocity errors, but with
the achieved positional accuracy these errors are no larger than $1.2
\kms$. 

The reduced and calibrated data set contains 19,200 STIS spectra, each
for a different aperture position in M15. We developed an algorithm
that co-adds the spectra for individual apertures, so as to extract
spectra of individual stars with minimum blending and maximum $S/N$.
Our data allowed us to extract spectra with an average $S/N > 5.5$ per
pixel for a total of 131 stars, out of which 15 stars are within $1''$
from the cluster center, 39 stars are within $2''$, and 87 stars are
within $5''$. In Paper~II we show that reliable line-of-sight
velocities can be derived for 64 of these stars, and that this
provides enough new information to significantly improve our
understanding of the dynamics and structure of M15.


\acknowledgments
Support for proposals \#8262 was provided by NASA through a grant from
the Space Telescope Science Institute, which is operated by the
Association of Universities for Research in Astronomy, Inc., under
NASA contract NAS 5-26555. We thank Pierre Dubath for helpful advice
in the early stages of this project, and Brian Yanny, Don Schneider
and John Bahcall for collaboration on the acquisition of the WFPC2
data. We thank the anonymous referee for useful feedback that helped
improve the presentation of the paper. 

\clearpage





\ifsubmode\else
\baselineskip=10pt
\fi


\clearpage


\ifsubmode\else
\baselineskip=14pt
\fi


\newcommand{\figcapcatimage}{Spatial distribution of the stars in the 
stellar catalog of M15. North is to the top and East is to the left.
All the stars in the catalog are shown; brighter stars are indicated
with slightly larger symbols. The footprints of the WFPC2 field of
view and of the individual CCDs are clearly visible. The cluster
center determined by G96 is at the origin. The cross marks the star
that was used to perform target acquisitions with
STIS.\label{f:catimage}}

\newcommand{\figcapCMD}{CMD of $V$ vs.~$B-V$ for the stars in the
stellar catalog of M15. The cross near the tip of the red giant branch
marks the star that was used to perform target acquisitions with
STIS.\label{f:CMD}}

\newcommand{\figcapHDslit}{Schematic illustration of the setup for the
spectroscopic observations of the calibration star HD 122563. The STIS
$X$ direction is perpendicular to the slit, and the $Y$ direction is
along the slit. Note that the three panels are offset from each other
by $2''$ in $Y$. The grating disperses the light that falls through
the slit such that wavelength increases in the positive $X$ direction.
The origin of the coordinate system is defined as the position of the
star after the target acquisition procedure. The symbols denote the 14
different positions relative to this origin at which the star was
subsequently placed using small telescope slews. At each position two
exposures were taken which were co-added during data reduction. The
bold dotted lines indicate the positions of the slit and the CCD
pixels along the slit, as determined from the data (see
Section~\ref{ss:psflsf}). Note that the slit was not centered on the
line $X=0$, i.e., the star was not perfectly centered in the slit
after the target acquisition procedure.\label{f:HDslit}}
 
\newcommand{\figcapundersamp}{HST/STIS spectra of the calibration star 
HD 122563. Top curve: a rectified 2D spectrum summed along all rows.
Bottom curve: the brightest row of the same rectified 2D spectrum.
Both spectra were normalized before plotting; for clarity, the top
spectrum was offset vertically by $1$ unit. Due to spatial
undersampling, the individual rows of 2D rectified STIS spectra show
low-frequency undulations multiplied into the spectrum. Due to their
low-frequency, these undulations do not compromise our ability to
measure radial velocities using cross-correlation techniques (see
Figure~\ref{f:crosscor}).\label{f:undersamp}}

\newcommand{\figcapcrosscor}{The cross-correlation function $f_{\rm cc}$ 
of the two spectra shown in Figure~\ref{f:undersamp}, as function of
the applied shift $\Delta V$ (in $\kms$). The bottom curve shows the
result without any continuum-subtraction or low-frequency
filtering. There is a narrow peak at $\Delta V = 0$ (the correct
value), but there are secondary peaks of almost equal amplitude due to
the low-frequency undulations in one of the two spectra. The top curve
shows the result when continuum-subtraction and low-frequency
filtering are applied as recommended in cross-correlation techniques
(Tonry \& Davies 1979); for clarity, the top curve was offset
vertically by $1$ unit. The peak at $\Delta V = 0$ is now
unmistakable. This illustrates that velocity determinations are not
adversely affected by the low-frequency undulations in the observed
spectra.\label{f:crosscor}}

\newcommand{\figcapcalibtot}{Analysis of the calibration star 
observations. Each data point corresponds to a fixed position of the
star with respect to the slit. The quantity $X$ along the abscissa is
the position of the star in the direction perpendicular to the slit.
The symbols and coordinate system are the same as in
Figure~\ref{f:HDslit}. A single grand-total one-dimensional spectrum
was constructed for each position of the calibration star, by
co-adding all rows in the reduced two-dimensional frame. The quantity
${\cal C}$ shown in the top panel is the count-rate in the grand-total
spectrum, in arbitrary units. The quantity ${\cal S}$ shown in the
bottom panel is the wavelength shift ${\cal S}$ (in pixels) of the
grand-total spectrum with respect to a fixed arbitrary standard. The
solid curves show the fit to these data obtained with the
multi-Gaussian PSF model described in Section~\ref{ss:psflsf}. The
vertical dotted line shows the inferred value $X_{\rm slit}$ of the
slit center (see also Figure~\ref{f:HDslit}). The horizontal dotted
line is the model value ${\cal S} = {\cal S}_{\rm slit}$ at the slit
center.\label{f:calibtot}}

\newcommand{\figcapPSF}{Heavy dotted curves are the results of calculations 
with the TinyTim software package for: {\bf (Top panel)} $E(r)$, the
encircled flux curve; {\bf (Middle panel)} ${\rm PSF}(r)$ on a linear
scale; and {\bf (Bottom panel)} ${\rm PSF}(r)$ on a logarithmic
scale. The thin solid curve shows the fit to these calculations
obtained with the multi-Gaussian PSF model described in
Section~\ref{ss:psflsf}.\label{f:PSF}}

\newcommand{\figcapcenter}{Schematic diagram of the central 
$2.5'' \times 2.5''$ of M15. The orientation and coordinate system are
as in Figure~\ref{f:catimage}. The origin is at the position of the
cluster center determined by G96, which has a 1-$\sigma$ uncertainty
of $0.2''$ in each coordinate. The corresponding two-dimensional
$68.3$\% and $95.4$\% confidence regions on the cluster center are
shown as a solid and a dotted circle, respectively. All the stars in
the WFPC2 catalog that are brighter than $V=20$ are shown; brighter
stars are indicated with slightly larger symbols. HST/STIS
observations were obtained at 18 different slit positions with a slit
width of $0.095''$. The slit positions are displayed as they were
commanded to the telescope. The actual achieved slit positionings were
slightly different, as discussed in
Section~\ref{ss:slitpos}.\label{f:center}}

\newcommand{\figcapintprof}{The intensity profile 
along the slit for one of the observations. The solid curve shows the
flux observed in each row of the calibrated and rectified
two-dimensional frame, averaged over the full wavelength range of the
spectrum. Each row corresponds to $0.05071''$. The dashed curve shows
the predictions of the best fit model, obtained as described in
Section~\ref{ss:slitpos}. The model is based on the WFPC2 stellar
catalog described in Section~\ref{s:imaging} and the PSF model
described in Section~\ref{ss:psflsf}. It also includes a smooth model
for the light contribution of faint stars that are missing from the
WFPC2 catalog due to incompleteness (dotted curve). [For reference,
the intensity profile corresponds to the third slit from the right in
Figure~\ref{f:apertures}. Rownumbers increase towards the top left in
that figure; the peak at row 930 is due to star \#7313 in the WFPC2
Catalog, at position $(x,y) = (2.021'',-2.193'')$.].\label{f:intprof}}
 
\newcommand{\figcappointings}{Positional differences $(\Delta X,\Delta Y)$ 
between the expected pointing of the telescope (as commanded during
scheduling), and the actual pointing of the telescope (determined from
intensity profile modeling as in Figure~\ref{f:intprof}). The 18
symbols correspond to the 18 slit positions shown in
Figure~\ref{f:center}. Observations indicated with the same plot
symbol were obtained in the same visit (Visit~1: filled circles;
Visit~2: open circles; Visit~3: open squares; Visit~4: open triangles;
Visit~5: filled squares; and Visit~6: asterisks). The STIS $X$
direction is perpendicular to the slit, and the $Y$ direction is along
the slit.\label{f:pointings}}

\newcommand{\figcapapertures}{Schematic diagram of the central
$2.5'' \times 2.5''$ of M15, as in Figure~\ref{f:center}. The
indicated slit and pixel positions are those that were actually
achieved during the HST/STIS observations. The positions were
calculated from the observed intensity profiles along the slits (see
Section~\ref{ss:slitpos}), and are accurate to better than
$0.01''$. The coverage of the central region of M15 was not quite as
homogeneous as planned (compare to Figure~\ref{f:center}) due to small
pointing errors (see Figure~\ref{f:pointings}).\label{f:apertures}}

\newcommand{\figcapspectra}{Four HST/STIS spectra of M15 giants in our 
final sample of 131 stars. The IDs of the stars in our WFPC2 Catalog
are, from top to bottom: \#6290 (magnitude $V=13.74$, color $B-V =
1.01$, average $S/N = 126.1$ per pixel); \#6833 ($V=15.88$, $B-V =
0.85$, $S/N = 43.3$); \#8396 ($V=17.31$, $B-V = 0.68$, $S/N = 18.7$);
and \#10589 ($V=18.10$, $B-V = 0.64$, $S/N = 7.3$). Although the
bottom spectrum is quite noisy, the Mg b triplet lines at $\sim
5170${\AA} are still clearly visible. All spectra were normalized
before plotting. For clarity, the top three spectra were offset
vertically by $1$, $1.5$ and $2$ units,
respectively.\label{f:spectra}}


\ifsubmode
\figcaption{\figcapcatimage}
\figcaption{\figcapCMD}
\figcaption{\figcapHDslit}
\figcaption{\figcapundersamp}
\figcaption{\figcapcrosscor}
\figcaption{\figcapcalibtot}
\figcaption{\figcapPSF}
\figcaption{\figcapcenter}
\figcaption{\figcapintprof}
\figcaption{\figcappointings}
\figcaption{\figcapapertures}
\figcaption{\figcapspectra}
\clearpage
\else\printfigtrue\fi

\ifprintfig


\clearpage
\begin{figure}
\epsfxsize=\hsize
\centerline{\epsfbox{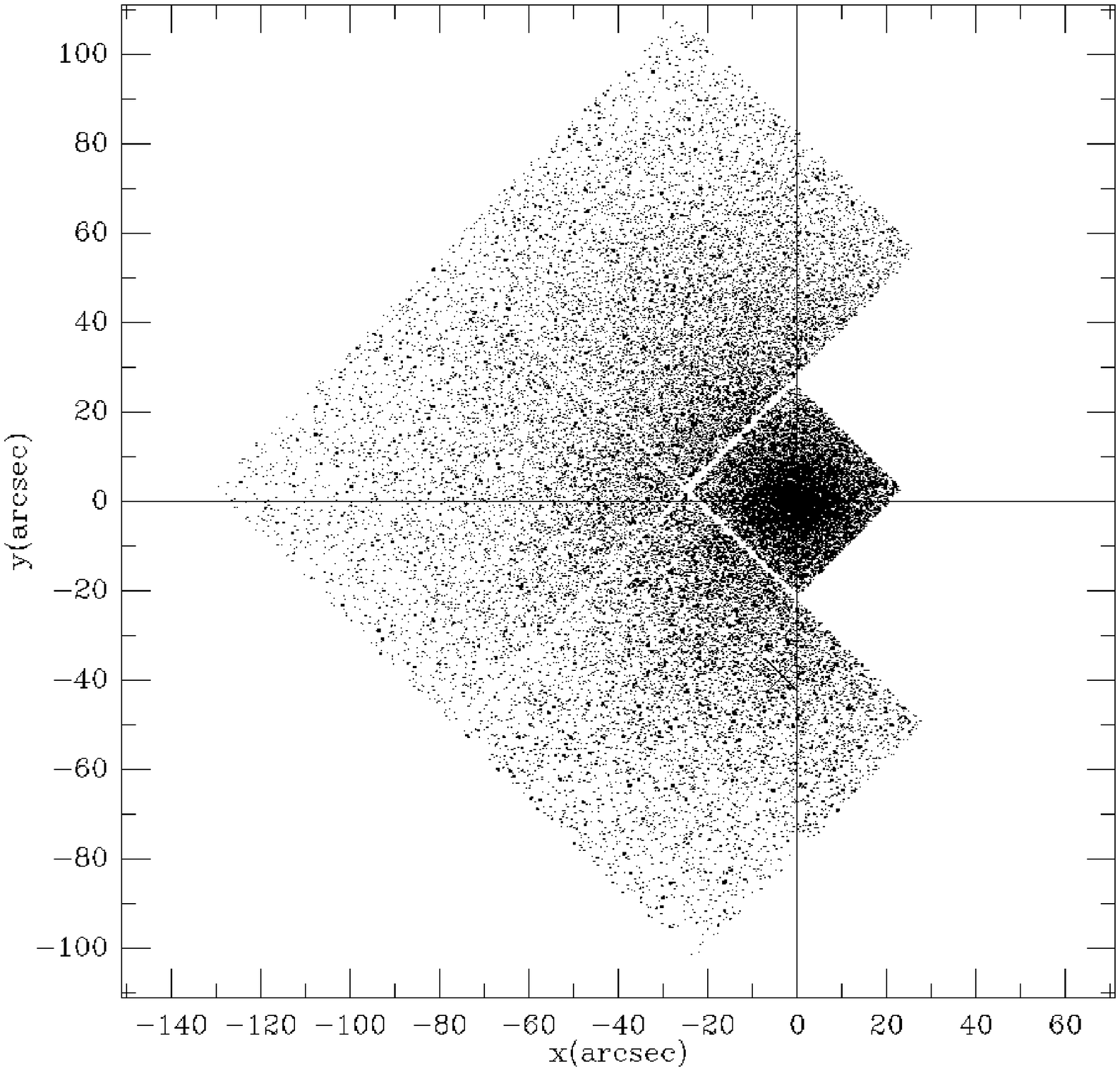}}
\ifsubmode
\vskip3.0truecm
\setcounter{figure}{0}
\addtocounter{figure}{1}
\centerline{Figure~\thefigure}
\else\figcaption{\figcapcatimage}\fi
\end{figure}


\clearpage
\begin{figure}
\epsfxsize=\hsize
\centerline{\epsfbox{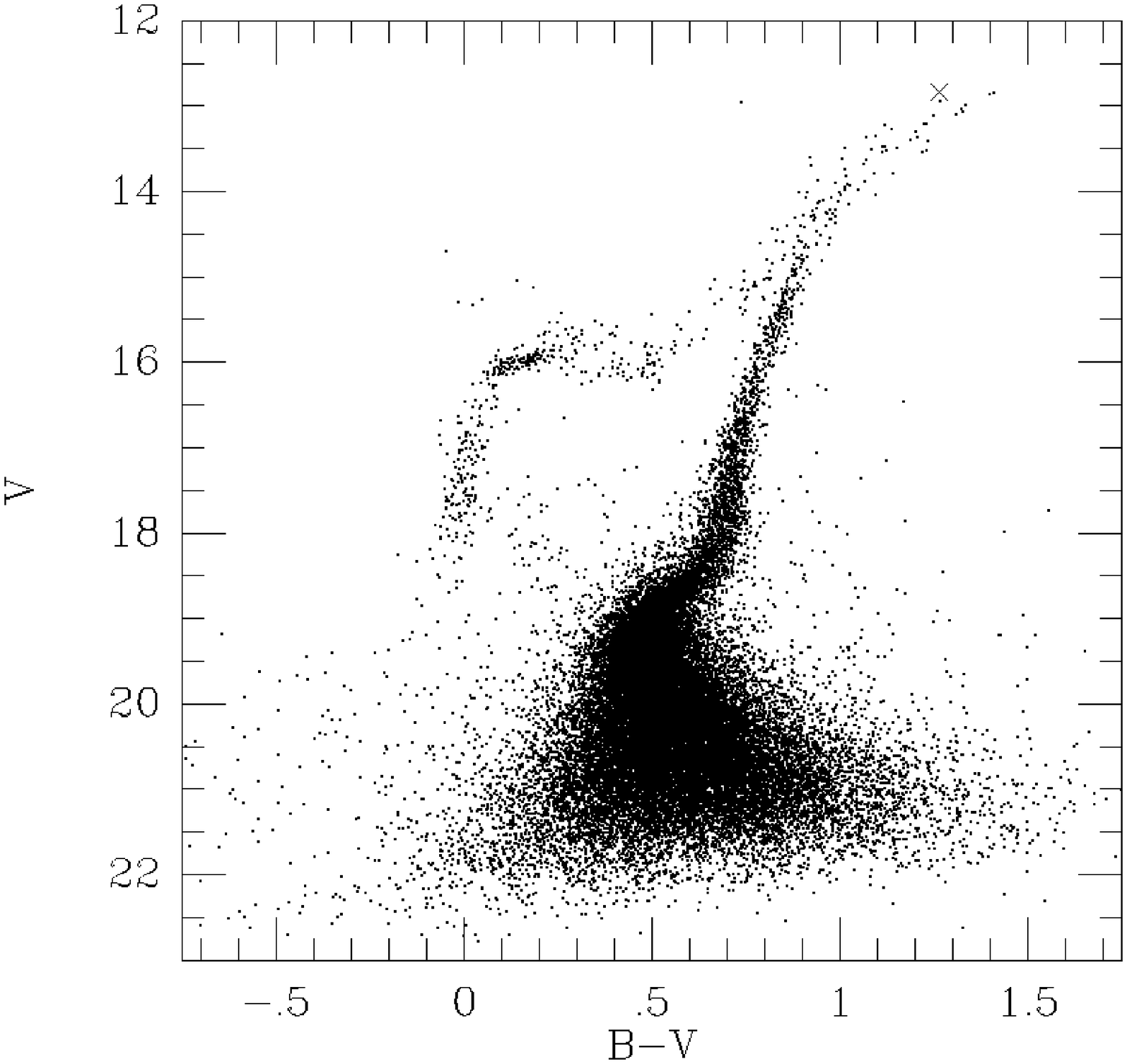}}
\ifsubmode
\vskip3.0truecm
\addtocounter{figure}{1}
\centerline{Figure~\thefigure}
\else\figcaption{\figcapCMD}\fi
\end{figure}


\clearpage
\begin{figure}
\epsfysize=15.0truecm
\centerline{\epsfbox{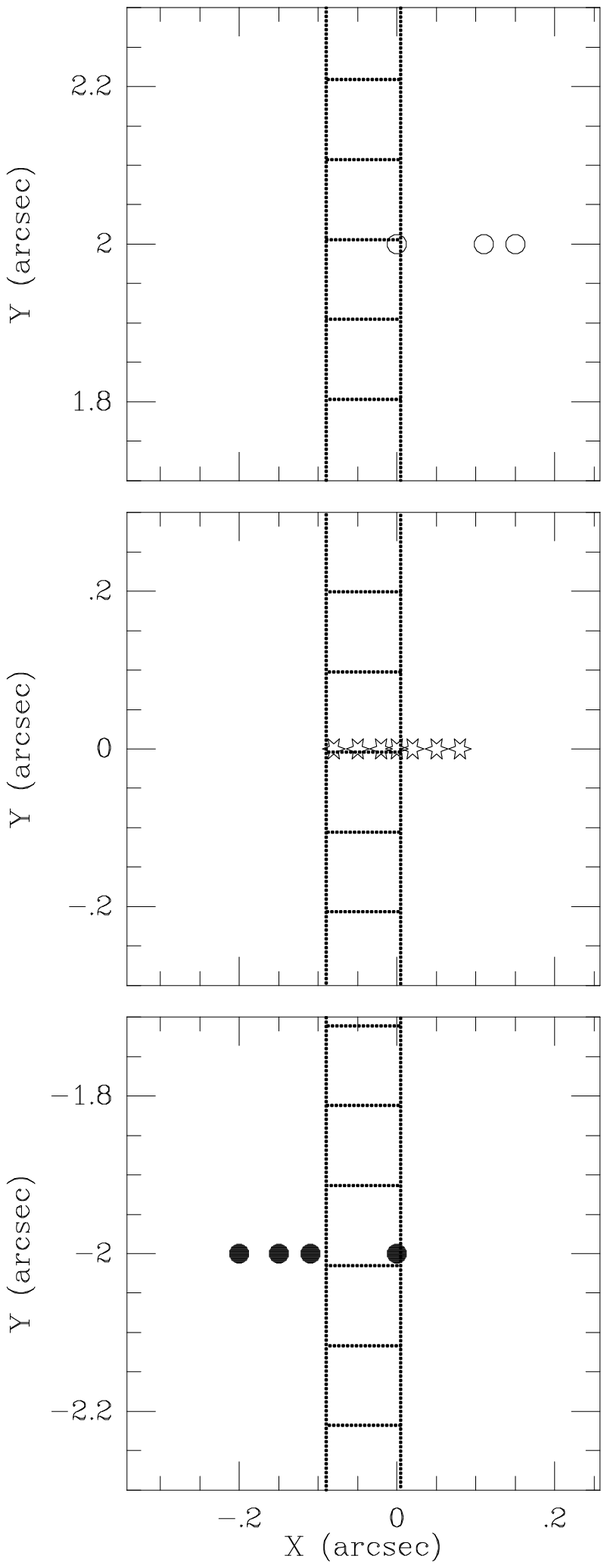}}
\ifsubmode
\vskip3.0truecm
\addtocounter{figure}{1}
\centerline{Figure~\thefigure}
\else\figcaption{\figcapHDslit}\fi
\end{figure}


\clearpage
\begin{figure}
\centerline{\epsfbox{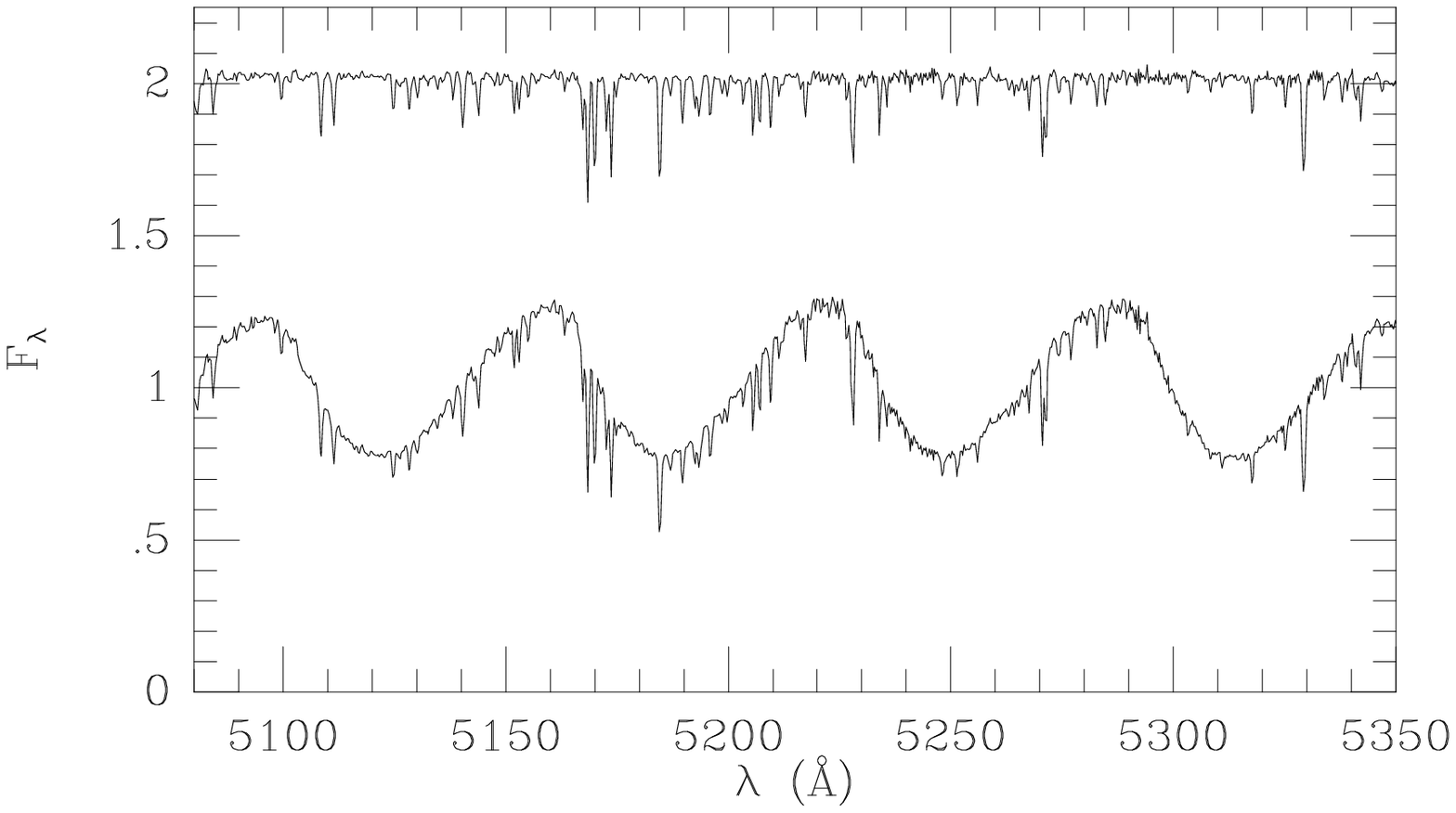}}
\ifsubmode
\vskip3.0truecm
\addtocounter{figure}{1}
\centerline{Figure~\thefigure}
\else\figcaption{\figcapundersamp}\fi
\end{figure}


\clearpage
\begin{figure}
\centerline{\epsfbox{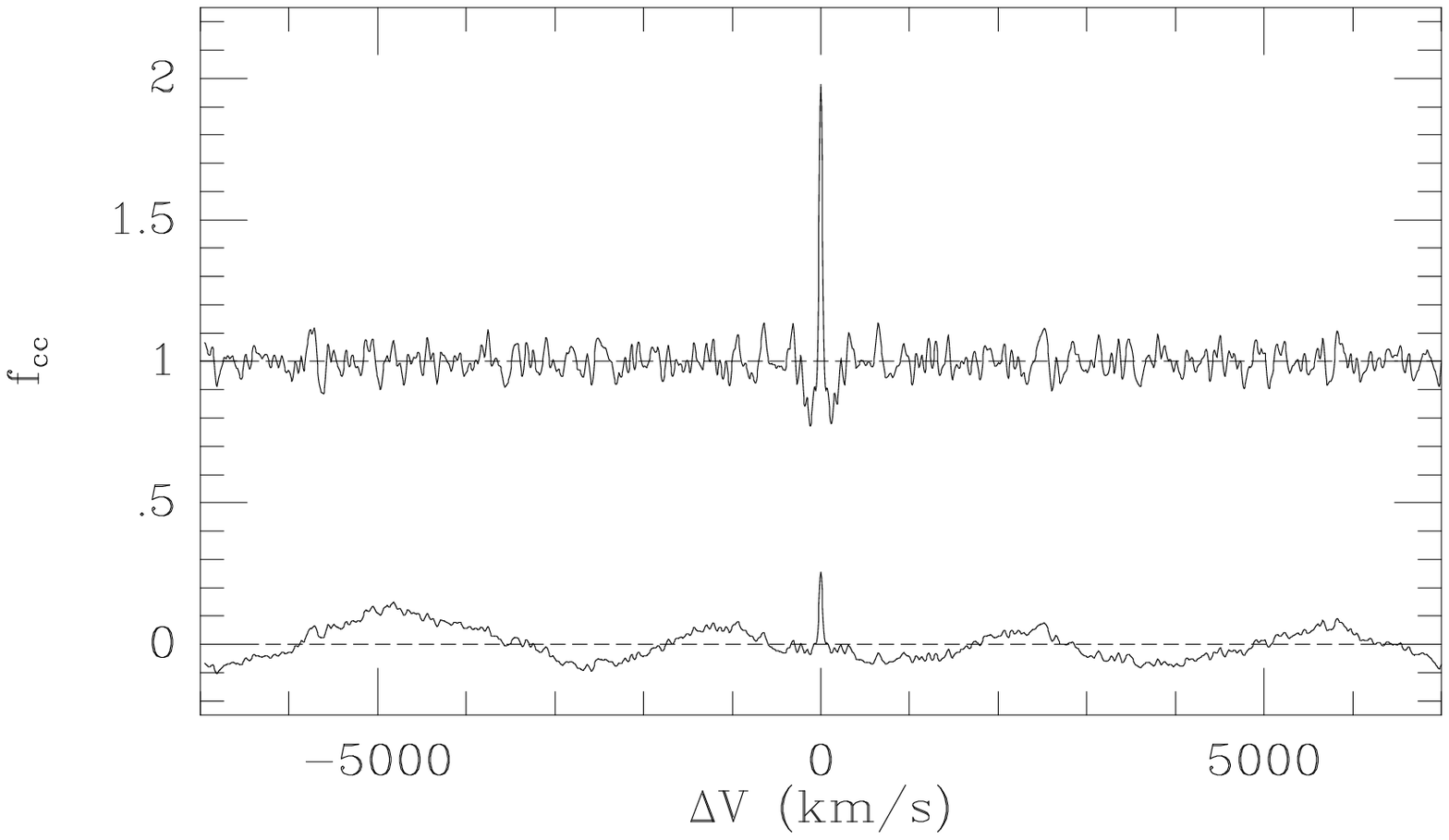}}
\ifsubmode
\vskip3.0truecm
\addtocounter{figure}{1}
\centerline{Figure~\thefigure}
\else\figcaption{\figcapcrosscor}\fi
\end{figure}


\clearpage
\begin{figure}
\epsfxsize=0.8\hsize
\centerline{\epsfbox{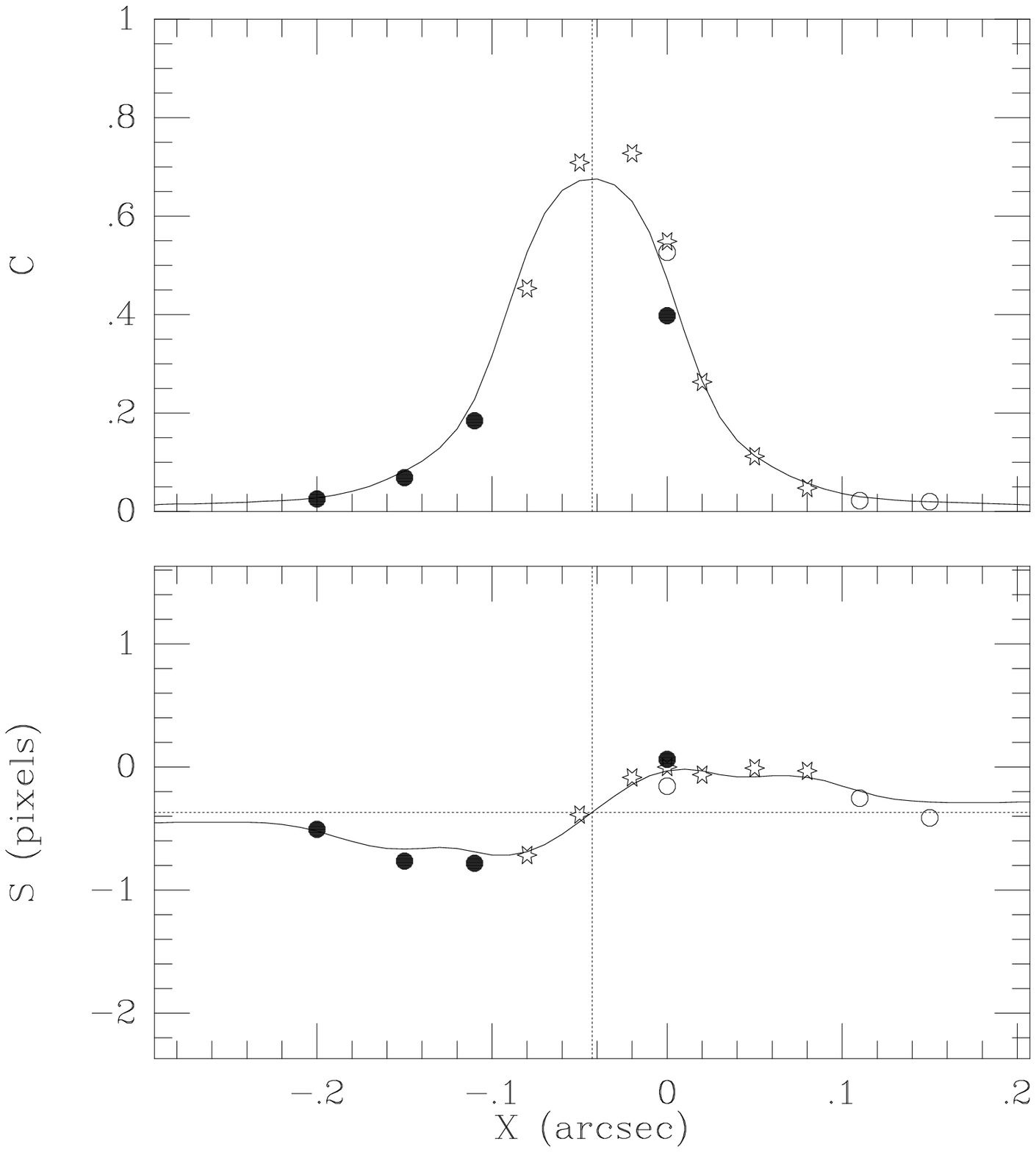}}
\ifsubmode
\vskip3.0truecm
\addtocounter{figure}{1}
\centerline{Figure~\thefigure}
\else\figcaption{\figcapcalibtot}\fi
\end{figure}


\clearpage
\begin{figure}
\centerline{\epsfbox{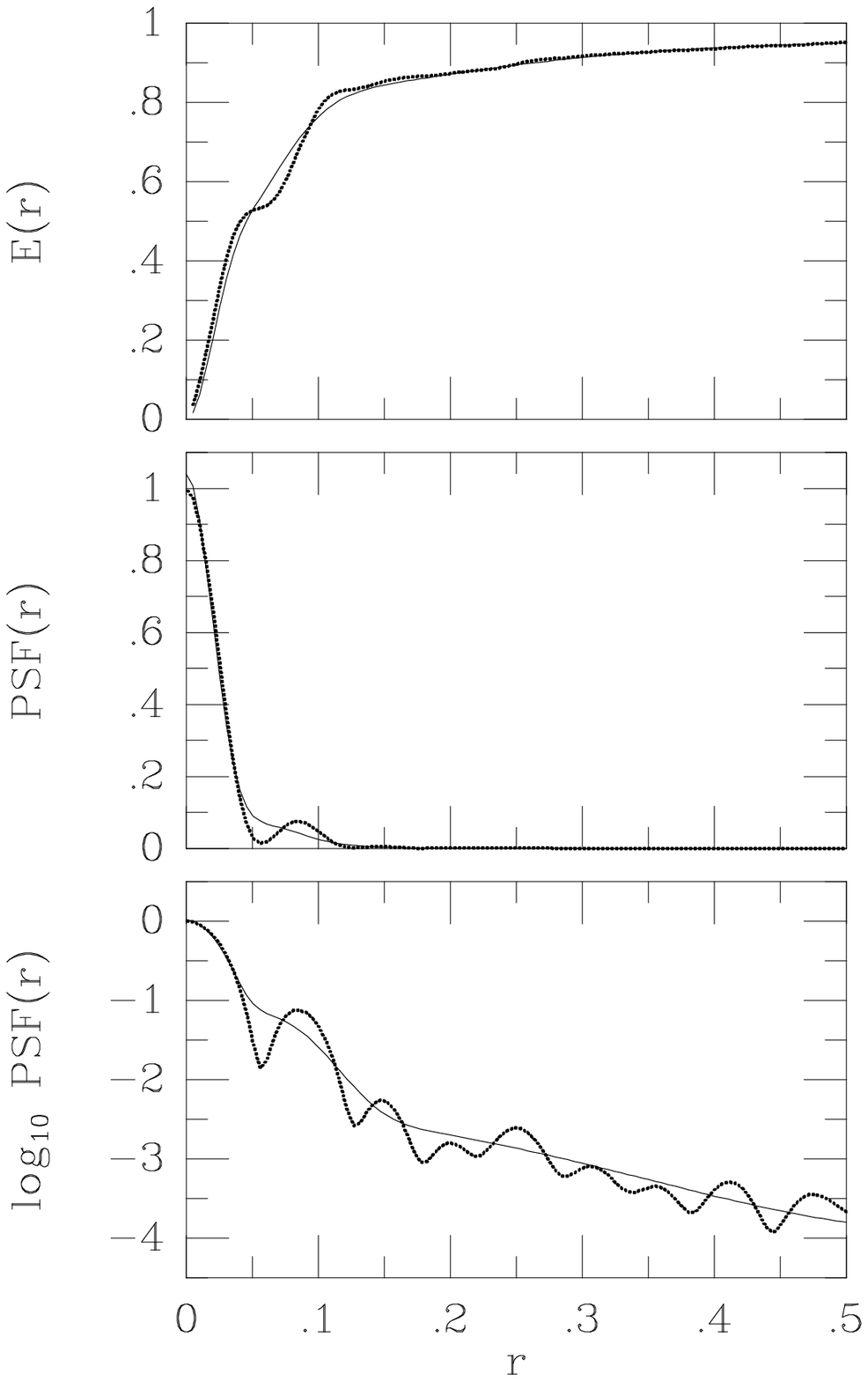}}
\ifsubmode
\vskip3.0truecm
\addtocounter{figure}{1}
\centerline{Figure~\thefigure}
\else\figcaption{\figcapPSF}\fi
\end{figure}


\clearpage
\begin{figure}
\centerline{\epsfbox{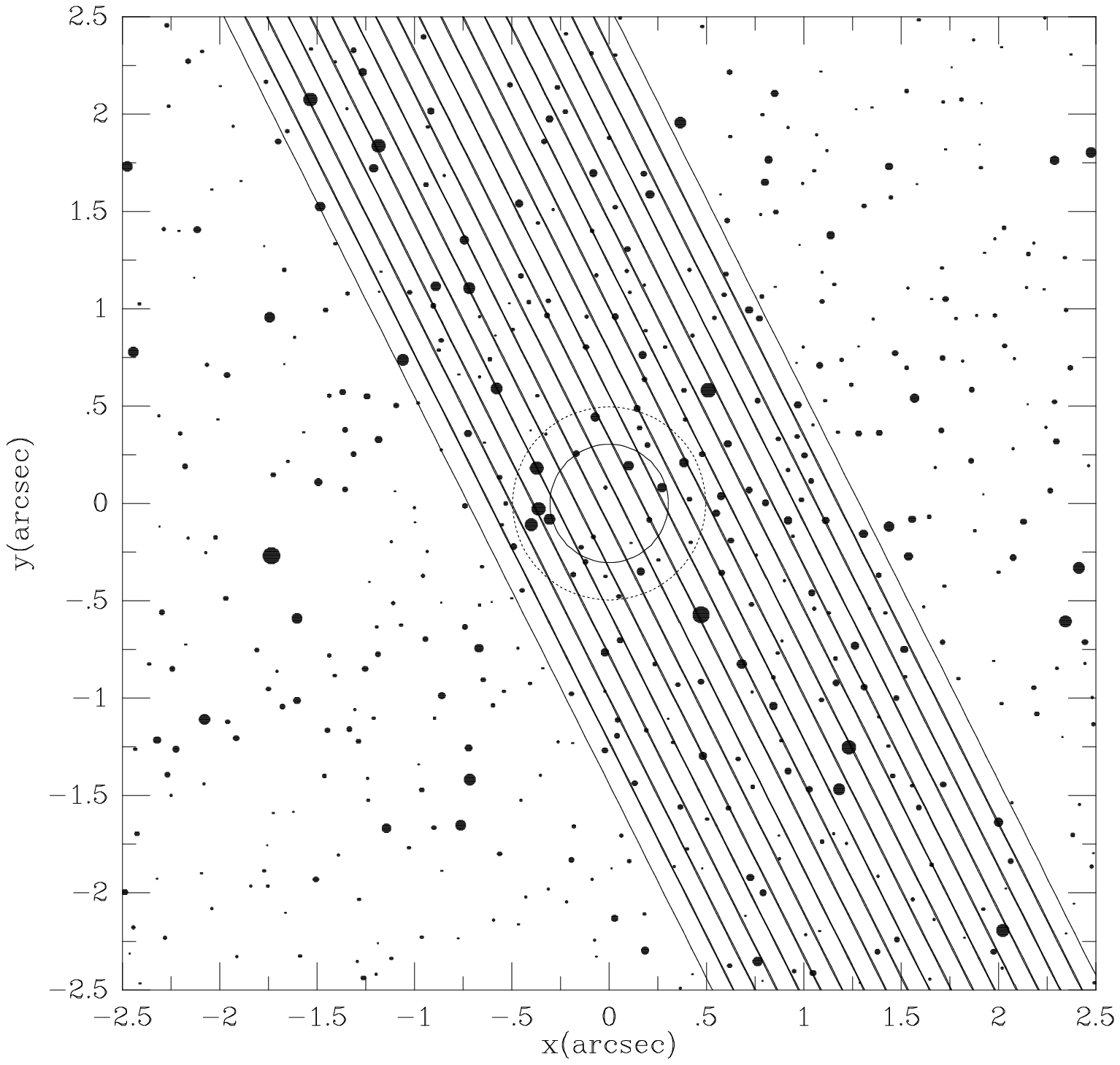}}
\ifsubmode
\vskip3.0truecm
\addtocounter{figure}{1}
\centerline{Figure~\thefigure}
\else\figcaption{\figcapcenter}\fi
\end{figure}


\clearpage
\begin{figure}
\epsfxsize=0.8\hsize
\centerline{\epsfbox{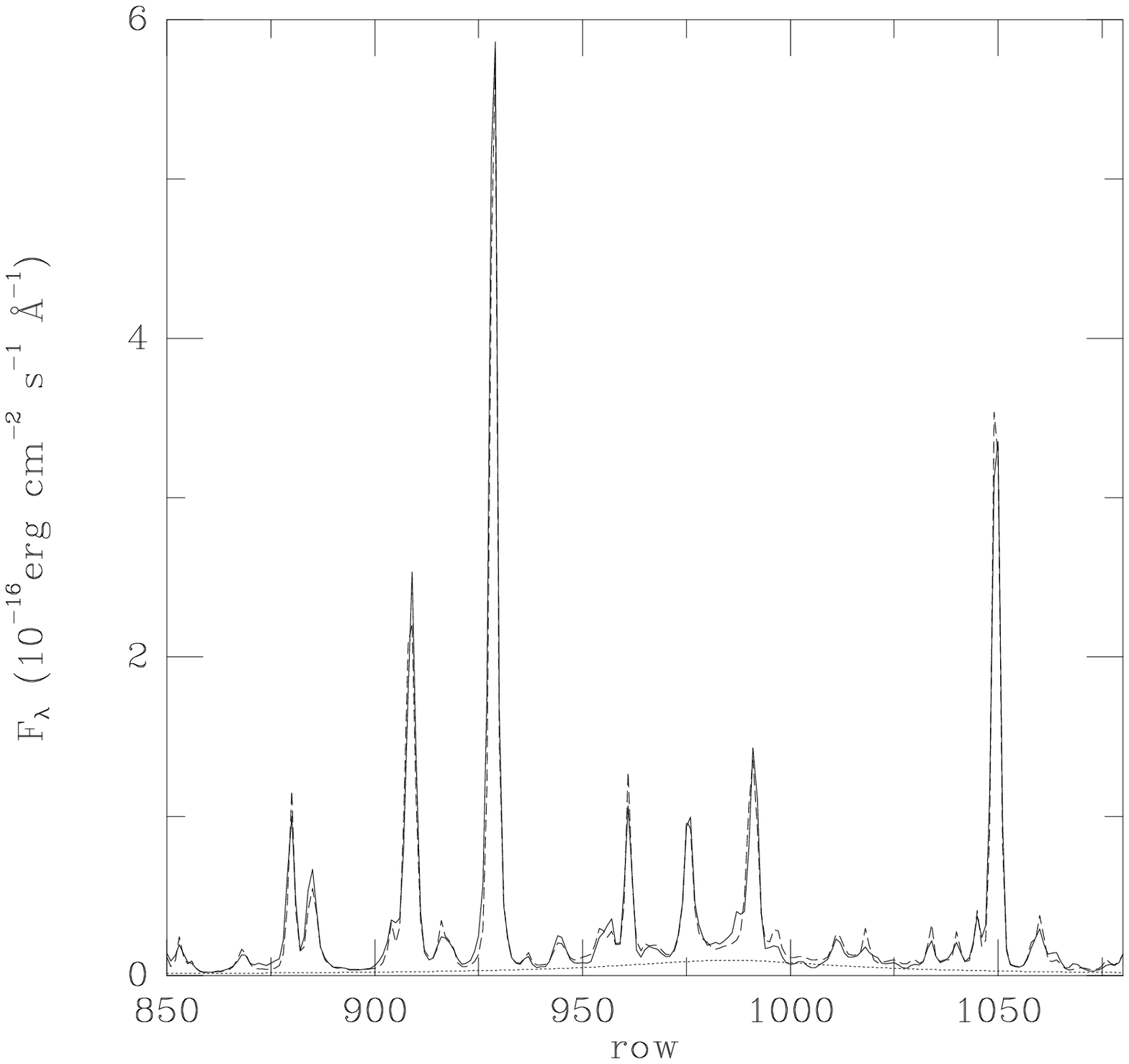}}
\ifsubmode
\vskip3.0truecm
\addtocounter{figure}{1}
\centerline{Figure~\thefigure}
\else\figcaption{\figcapintprof}\fi
\end{figure}


\clearpage
\begin{figure}
\centerline{\epsfbox{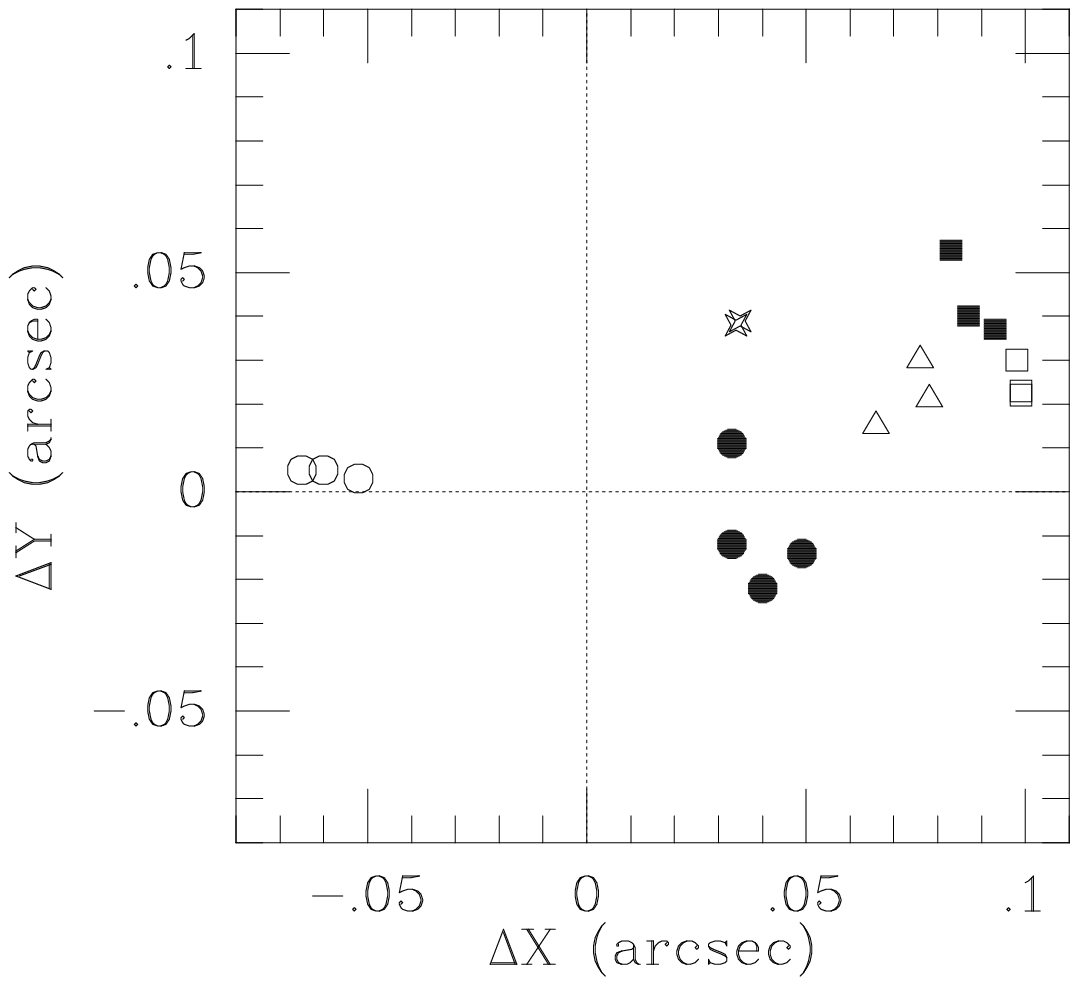}}
\ifsubmode
\vskip3.0truecm
\addtocounter{figure}{1}
\centerline{Figure~\thefigure}
\else\figcaption{\figcappointings}\fi
\end{figure}


\clearpage
\begin{figure}
\centerline{\epsfbox{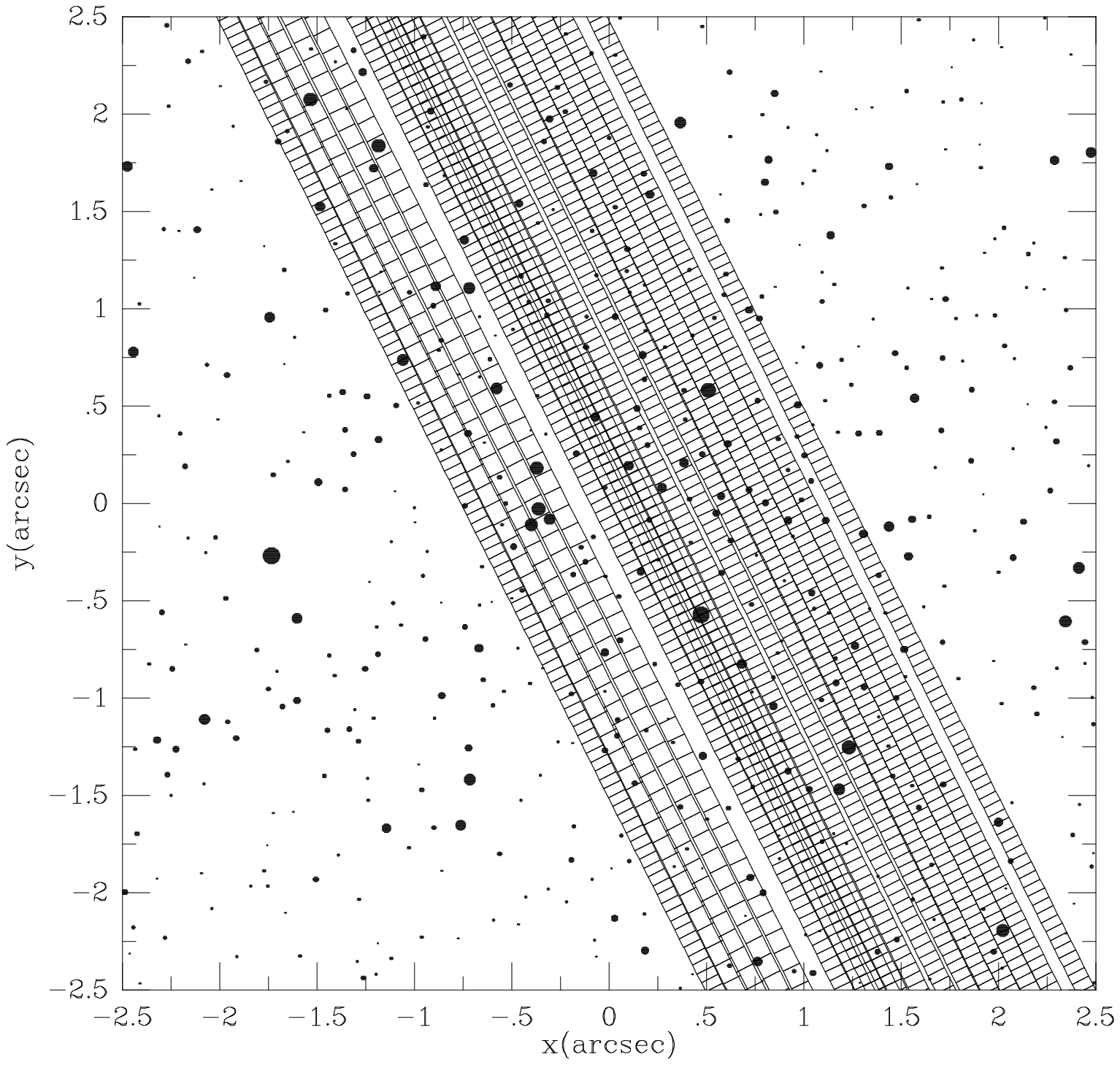}}
\ifsubmode
\vskip3.0truecm
\addtocounter{figure}{1}
\centerline{Figure~\thefigure}
\else\figcaption{\figcapapertures}\fi
\end{figure}


\clearpage
\begin{figure}
\centerline{\epsfbox{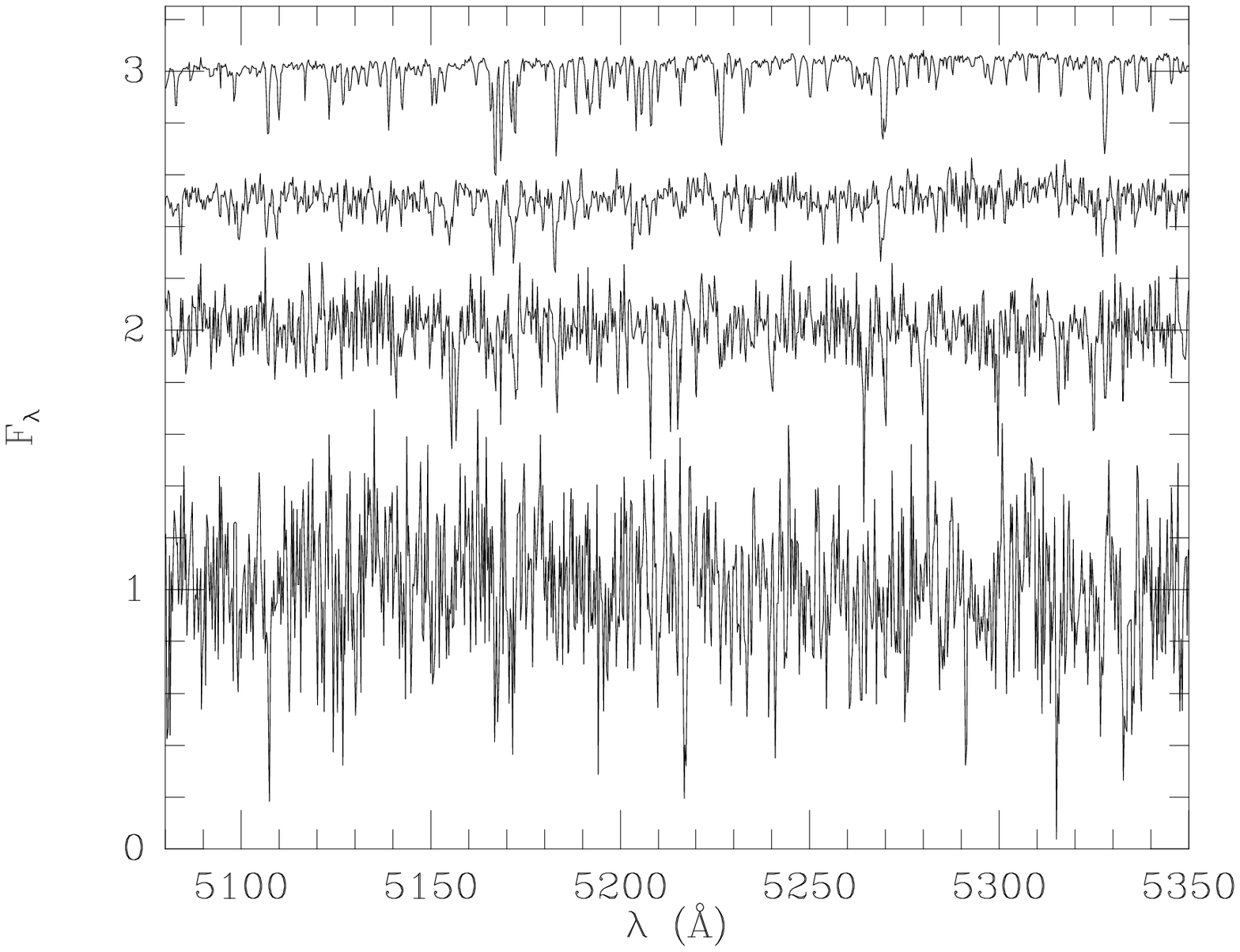}}
\ifsubmode
\vskip3.0truecm
\addtocounter{figure}{1}
\centerline{Figure~\thefigure}
\else\figcaption{\figcapspectra}\fi
\end{figure}


\fi


\clearpage
\ifsubmode\pagestyle{empty}\fi


\begin{deluxetable}{rrrccc}
\tablecaption{HST/WFPC2 Photometric Catalog of M15\label{t:catalog}} 
\tablehead{
\colhead{ID}  & \colhead{$\Delta {\rm RA}$} & 
\colhead{$\Delta {\rm DEC}$} & $V$ & $B$ & $U$ \\
 & \colhead{(arcsec)} & \colhead{(arcsec)} & & & \\
\colhead{(1)} & \colhead{(2)} & \colhead{(3)} & \colhead{(4)} & 
\colhead{(5)} & \colhead{(6)} \\
}
\startdata
    1 & 16.343 & 10.269 & 20.467 & 21.313 & 20.869 \\
    2 & 18.180 &  8.384 & 20.889 & 21.701 & 21.624 \\
    3 &  7.251 & 19.504 & 20.271 & 20.900 & 20.623 \\
    4 & 18.725 &  7.807 & 19.363 & 19.893 & 19.738 \\
    5 &  2.529 & 24.193 & 20.256 & 20.583 & 20.534 \\
    6 &  2.170 & 24.551 & 20.325 & 20.994 & 20.738 \\
    7 & 17.248 &  9.263 & 20.738 & 21.234 & 21.206 \\
    8 & 11.883 & 14.729 & 14.659 & 15.562 & 15.894 \\
    9 &  1.843 & 24.853 & 20.619 & 21.485 & 21.278 \\
   10 & 15.338 & 11.195 & 19.454 & 19.935 & 19.840 \\
   11 & 22.580 &  3.733 & 20.957 & 21.767 & 21.958 \\
   12 & 19.811 &  6.584 & 18.549 & 19.215 & 19.148 \\
   13 &  6.533 & 20.103 & 19.973 & 20.440 & 20.372 \\
   14 &  3.334 & 23.280 & 20.557 & 21.253 & 20.911 \\
   15 & 16.777 &  9.611 & 20.453 & 21.105 & 20.926 \\
   16 &  3.759 & 22.784 & 20.885 & 21.633 & 21.377 \\
   17 &  8.709 & 17.789 & 20.776 & 21.377 & 21.155 \\
   18 & 13.186 & 13.240 & 20.985 & 22.073 & 21.703 \\
   19 &  8.817 & 17.666 & 19.188 & 19.698 & 19.478 \\
   20 & 21.074 &  5.129 & 19.719 & 20.445 & 20.145 \\
   21 &  4.649 & 21.815 & 20.424 & 20.885 & 20.718 \\
   22 &  6.399 & 20.045 & 20.568 & 21.132 & 20.988 \\
   23 & 20.324 &  5.839 & 20.359 & 21.056 & 20.829 \\
   24 &  6.592 & 19.834 & 19.828 & 20.382 & 20.294 \\
   25 & 16.308 &  9.954 & 20.307 & 21.043 & 20.993 \\
\enddata
\tablecomments{First 25 entries of the HST/WFPC2 photometric catalog described
in Section~\ref{ss:photometric}. The full catalog contains 31,983
stars and is distributed electronically as part of this
paper. Column~(1) is the ID number of the star. Columns~(2) and~(3)
give positions $({\rm RA}, {\rm DEC})$ of each star, measured with
respect to the M15 cluster center as described in
Section~\ref{ss:astrometric}. Columns~(4), (5) and~(6) give the
broad-band $V$, $B$ and $U$ magnitudes of the stars.}
\end{deluxetable}


\begin{deluxetable}{ccc}
\tablecaption{Multi-Gaussian PSF model parameters\label{t:psf}}
\tablehead{
\colhead{$i$}  & \colhead{$\gamma_i$} & \colhead{$\sigma_i$} \\
               &                      & \colhead{(arcsec)}   \\
}
\startdata
1 & $ 0.589$ & $0.020$ \\
2 & $-0.625$ & $0.032$ \\
3 & $ 0.836$ & $0.044$ \\
4 & $ 0.129$ & $0.165$ \\
5 & $ 0.071$ & $0.572$ \\
\enddata
\tablecomments{The listed parameters describe the multi-Gaussian 
PSF model, defined by equation~(\ref{PSFGauss}), that was used for the
analysis and interpretation of the M15 spectroscopy. The PSF model
provides the optimum fit to the combined data shown in
Figures~\ref{f:calibtot} and~\ref{f:PSF}, as described in
Section~\ref{ss:psflsf}.}
\end{deluxetable}


\begin{deluxetable}{ccccc}
\tablecaption{M15 STIS Observations: Observation Log\label{t:obslog}}
\tablehead{
\colhead{position} & \colhead{$T_{\rm exp}$} &
\colhead{$N_{\rm exp}$} & \colhead{rebin} & \colhead{visit} \\
\colhead{(arcsec)} & \colhead{(sec)} & & & \\
\colhead{(1)} & \colhead{(2)} & \colhead{(3)} & \colhead{(4)} & \colhead{(5)} \\
}
\startdata
$+0.6$ & 2862 & 4 & 1 & 6 \\
$+0.5$ & 2790 & 4 & 2 & 1 \\
$+0.4$ & 2760 & 3 & 2 & 1 \\
$+0.3$ & 2760 & 3 & 2 & 1 \\
$+0.2$ & 2760 & 3 & 2 & 1 \\
$+0.1$ & 3113 & 4 & 1 & 2 \\
$+0.0$ & 3578 & 4 & 1 & 2 \\
$-0.1$ & 3556 & 4 & 1 & 2 \\
$-0.2$ & 3113 & 4 & 1 & 3 \\
$-0.3$ & 3578 & 4 & 1 & 3 \\
$-0.4$ & 3556 & 4 & 1 & 3 \\
$-0.5$ & 3113 & 4 & 1 & 4 \\
$-0.6$ & 3578 & 4 & 1 & 4 \\
$-0.7$ & 3556 & 4 & 1 & 4 \\
$-0.8$ & 3113 & 4 & 1 & 5 \\
$-0.9$ & 3578 & 4 & 1 & 5 \\
$-1.0$ & 3556 & 4 & 1 & 5 \\
$-1.1$ & 3554 & 4 & 1 & 6 \\
\enddata
\tablecomments{All observations were taken with the {\tt
52X0.1} slit along position angle ${\rm PA} = 26.65^{\circ}$ on the
sky, measured from North over East. Column~(1) lists the offset of the
slit from the M15 center, measured in the direction of position angle
PA=$116.65^{\circ}$. The listed offset is the one that was commanded
to the telescope; the actual offsets as determined from the data are
slightly different, as described in Section~\ref{ss:slitpos}. The
position of the M15 center was chosen as described in
Section~\ref{ss:astrometric}.  Column~(2) lists the total exposure
time at the given slit position. Column~(3) lists the number of
exposures. Column~(4) lists the number of spatial pixels that were
rebinned `on-chip' during CCD read-out. Column~(5) lists the number of
the visit in which the observations were done.}
\end{deluxetable}



\end{document}